\documentclass[fleqn,usenatbib]{mnras}
\usepackage{natbib}
\usepackage{hyperref}
\usepackage{svg}

\newcommand{\microHz}{{$\mu$Hz}}
\usepackage[utf8]{inputenc}
\usepackage{graphicx}

\title[Low-degree solar rotational splitting]{Low-degree solar rotational splitting from 45 years of BiSON observations}
\author[R. Howe et al.]{
Rachel Howe,$^{1}$\thanks{E-mail: r.howe@bham.ac.uk (RH)}
W.~J.~Chaplin,$^{1}$
Y.~P.~Elsworth,$^{1}$
S.~J.~Hale,$^{1}$
and M.~B.~Nielsen$^{1}$
\\
$^{1}$School of Physics and Astronomy, University of Birmingham, Edgbaston, Birmingham B15 2TT, UK} 

\date{Accepted XXX. Received YYY; in original form ZZZ}

\pubyear{2023}

\begin{document}
\label{firstpage}
\pagerange{\pageref{firstpage}--\pageref{lastpage}}
\maketitle

\begin{abstract}
We present solar low-degree rotational splitting values based on a new analysis of Sun-as-a-star observations from the Birmingham Solar Oscillations Network, covering a 16,425-day period from 1976 December 31\,--\,2021 December 20 with a duty cycle of 57 per cent.  The splitting values are estimated from the power spectrum using a Markov Chain Monte Carlo sampling method, and we also present for comparison the results from an analysis of 100 realizations of synthetic data with the same resolution and gap structure. Comparison of the scatter in the results from the synthetic realizations with their estimated uncertainties suggests that for this data set the formal uncertainty estimates are about 30 per cent too small. An upward bias in the splittings at frequencies above {2200\,\microHz}, where the components are not fully resolved, is seen in both the observed and synthetic data. When this bias is taken into account our results are consistent with a frequency-independent synodic rotational splitting value of 400\,nHz.

\end{abstract}

\begin{keywords}
Sun: helioseismology -- Sun:rotation
\end{keywords}

\section{Introduction}

The low-degree solar $p$-modes are one of the very few tools available to probe the structure and rotation of the deepest parts of the solar interior. In order to use the information they provide, we need both to determine their properties as accurately as possible, and to have realistic estimates of the uncertainties on these measurements. The Birmingham Solar Oscillations Network (BiSON) has monitored these oscillations since the mid-1970s, originally in short observing campaigns with one or more ground-based instruments and later from an automated six-site network, giving us the longest time series of such observations available to date. In this article we present solar rotational splitting values obtained by fitting a single Fourier power spectrum covering almost the entire lifetime of the project to date.

\begin{figure}
\includegraphics[width=\linewidth]{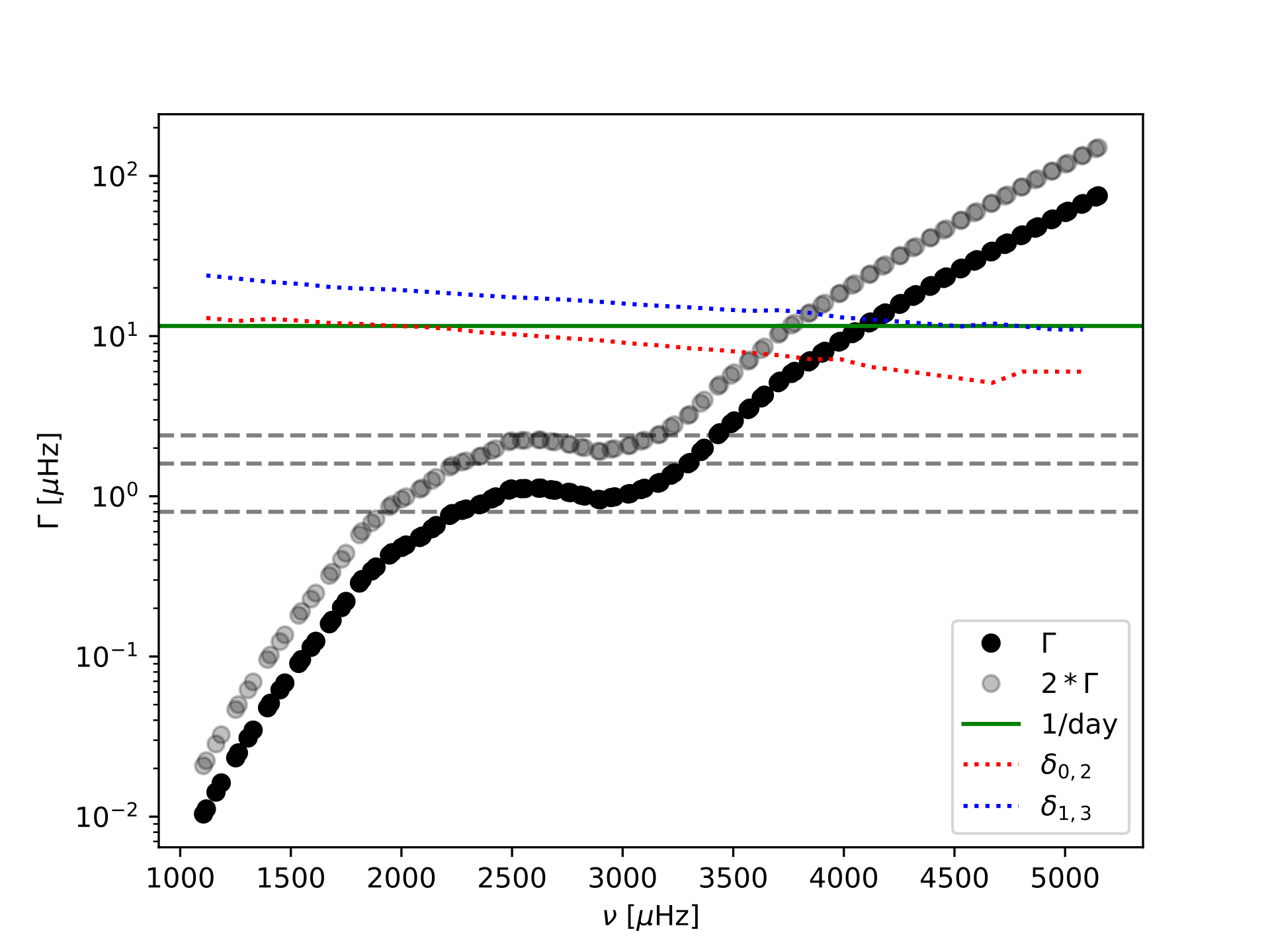}
  \caption{Mode linewidth for low-degree solar $p$-modes as a function of frequency. The dashed grey lines indicate the $m=\pm l$ splitting values for $l$=1, 2, and 3 modes; the dotted curves indicate the $l=2/0$ and $l=3/1$ separation. The solid horizontal line corresponds to the 1/day 
alias frequency.}
\label{fig:guesswidth}
\end{figure}

Solar rotation lifts the degeneracy between spherical harmonics with the same degree $l>0$ and different azimuthal order $m$, giving rise to a multiplet in which the variation of the frequency with $m$ contains information about the rotation and asphericity. 
The solar rotation period of approximately 27 days corresponds to an angular frequency of about 430\,nHz, but because the Earth orbits the Sun once a year, which translates to an angular frequency of 31.7\,nHz, the ``synodic'' rate that is measured from Earth-based observations is approximately 400\,nHz. This means that the difference in frequency between the $m=-l$ and $m=l$ components at a given degree $l$ and radial order $n$ is about $l\times 800$\,nHz. Figure~\ref{fig:guesswidth}, which is based on the first-guess table discussed below in Section~\ref{sec:optimization}, shows how the $m=\pm l$ splittings for $l$=1, 2, and 3 relate to the mode width; because peaks need to be separated by at least twice their FWHM $\Gamma$ to avoid a biased or unstable fit when using unconstrained Maximum Likelihood Estimation fitting \citep[see, for example][]{1998A&AS..131..539H}, we show both the $\Gamma$ and $2\Gamma$ values as a function of frequency. The $l=1$ components are separated by $2\Gamma$ only below about 1800\,\microHz, while the outer components of $l=3$ are $2\Gamma$ apart up to {3200\,\microHz}. The figure also shows the separation between frequencies for the $(l+2,n-1)/(l,n)$ mode pairs for $l=0$ and $l=1$; we can see that at frequencies above about {3600\,\microHz} for $l=2/0$ and {3800\,\microHz} for $l=3/1$ the mode pairs are not resolved, so it would be challenging if not impossible to estimate the splittings of individual modes in this regime to any meaningful precision.

BiSON data have been used to estimate low-degree rotational splittings since shortly after the deployment of the worldwide network. \cite{1995Natur.376..669E} used BiSON data from 1992 January to 1994 August, divided into 16-month segments, to measure the rotational splittings of $l=1$ modes between 1473 and {2559\,\microHz} (radial order $n$ from 9 to 17), of $l=2$ modes between 1536 and {2486\,\microHz} ($9 \leq n \leq 16$) , and of $l=3$ modes from 2138 to {2676\,\microHz} ($13 \leq n \leq 17$). \citet{2001MNRAS.327.1127C} analyzed an 8-yr set of BiSON data from 1992 January to 1999 December; they give splittings down to $l=1, n=9$ (1473\,\microHz), with the lowest-frequency $l=1$ splitting having an uncertainty of 5.3\,nHz.

\cite{2014MNRAS.439.2025D} obtained splittings and other mode parameters for modes at frequencies down to 1185 microHz ($n=7$), using a Bayesian algorithm to fit to a spectrum based on 22\,yr of BiSON data from 1991 January to 2012 December. They cite a ``duty cycle efficiency'' for this dataset of 68.3 per cent in a time series prepared to optimize for low-frequency noise.

\cite{2008SoPh..251..119G} analyzed a time series of 4182 days from 
1996 April 11 to 
2007 September 22 
from the Global Oscillations at Low Frequencies (GOLF) instrument onboard the Solar and Heliospheric Observatory, with 94 per cent duty cycle. They were able to measure splittings down to the $l=1,n=7$ (1185\,\microHz) mode.
They cite \cite{2006MNRAS.369..985C} on the possible bias of splittings by leakage from higher-degree modes (e.g. $l=4$
impinging on the $l=2$ fitting window), and hence they use a {45\,\microHz} window to fit each mode pair. They state that the $l=1$ splittings are
 ``roughly constant'' for the smaller fitting windows up to {3400\,\microHz} but increase at higher frequencies than that, whatever window is used, so they trust the splittings only up to {3400\,\microHz}.

\section{Data}
The main data set that we analyze here is based on nearly 45 years of data from BiSON, prepared as described by \citet{2014MNRAS.441.3009D}. The time series is zero-padded 
to give an integer multiple of 365 days and covers  $45\times365$=16,425 days from 1976 December 31 to 2021 December 20, with an overall duty cycle of 57 per cent. The performance of the network in its early years has been described by \cite{1996SoPh..168....1C}, and a more current overview was given by \cite{2016SoPh..291....1H}.

\section{Analysis}

We form the acoustic power spectrum by carrying out a Fast Fourier Transform of the prepared time series between the selected dates, in which any missing data are replaced by zeros in the detrended time series.

\subsection{Peak Profile}
We fit the spectrum using a model in which each mode or rotationally split component is 
an asymmetric Lorentzian function of the frequency $\nu$, described by the formula \label{eq:rh1}
\begin{equation}
P(\xi)=\left({h\over{1+\xi^2}}\right)\times[(1+2b\xi)],
\end{equation}
where
\begin{equation}
\xi=2(\nu-\nu_0)/\Gamma,
\end{equation}
$\nu_0$ is the frequency of the Lorentzian component, $\Gamma$ its
width,
$h$ its height, and $b$ is a fractional parameter characterizing the
asymmetry. The expression simplifies to the  normal Lorentzian for $b=0$.
This is the formulation proposed by \citet{1998ApJ...505L..51N}, but the quadratic term in $b$ is suppressed, as proposed by \citet{2009ApJ...694..144F}, in order to ensure that the value of the expression is small far from the central frequency.

\subsection{Rotational multiplets}    

Because rotational splitting lifts the degeneracy between modes of the same $l$ and different $m$, for each $(l,n)$ there are potentially $2l+1$ components of different $m$. As the inclination of the Sun's rotation axis to the observer is close to zero, we assume that components with $l-m$ odd have zero amplitude: \cite{2014MNRAS.439.2025D}, who also neglect these components, estimate their size at less than three per cent of the $|m|=l$ power. In practice, therefore, we only deal with $l+1$ components for a mode of degree $l$: $m=\pm 1$ for $l=1$, $m=0, m=\pm 2$ for $l=2$, and $m=\pm 1, m=\pm 3$ for $l=3$. Furthermore, we assume that the frequency and amplitude of components within a multiplet are symmetric about $m=0$, and that components $m,m^{\prime}$ within a multiplet are separated by $(m-m^{\prime})\delta\Omega$, where $\delta\Omega$ is our ``splitting'' parameter; following \citet{2014MNRAS.439.2025D}, as we are concerned only with low-degree modes we do not take into account differential rotation or asphericity. The asphericity term for $l=2$ is just detectable in BiSON data at epochs of high solar activity \citep{2003MNRAS.343..343C}, but while it might bias the frequency measurement it should not bias the splitting. For the $l=3$ multiplet, because of differential rotation, measuring the splitting in this way is not quite equivalent to measuring the first-order term of the polynomial expansion of frequency as a function of $m$, but the difference (which we estimate at around 5 \,nHz) will be within the uncertainties for all but the lowest-frequency modes, and for those the signal-to-noise ratio of the inner components is low, so it is not practical to fit the differential rotation term. The height of the $m=0$ component of the $l=2$ multiplet and the $m=\pm 1$ components for $l=3$ modes are scaled by a visibility factor $V_l$ relative to the $m=\pm l$ component; we base the prior distributions for these factors on values of 0.54 for $V_2$ and 0.38 for $V_3$. These values correspond to the ones used to construct synthetic BiSON-like data as described by  \cite{2006MNRAS.369..985C}, based on the fitting results of  \citet{2001MNRAS.327.1127C}. The asymmetry and width parameters are also assumed to be the same for all components within a multiplet. The full model is built up by adding the asymmetric Lorentzian functions for each component in an $l=2/0$ or $l=3/1$ pair to a flat background offset $c$.

\subsection{Window-function convolution}
Ground-based observations, even with a network, tend to have a daily periodicity in their observing window; the solar spectrum is convolved with the power spectrum of the window function \citep{1993A&A...280..704L}. This results, among other effects, in ``sidelobes'' at 1/day (11.57\,\microHz) on either side of each solar mode, which for some orders coincide with the other mode in an $(l,n)/(l-2,n+1)$ pair.  To mitigate this, as the last step of evaluating the model function we convolve the model with the power spectrum of the window function (a sequence of ones and zeros of the same length as the data, where a non-zero value corresponds to data being present), truncated at {$\pm 100$\,\microHz}. 
The convolution is implemented such that the model outside the fitting window is assumed to take the background offset value $c$, to avoid step-function effects in the convolved model that would occur if it were set to zero. To be specific, we use the {\tt {astropy.convolution.convolve\_fft}} function and set the {\tt{ fill\_value}} keyword to $c$.

\subsection{Fitting Window}
\begin{figure}
\includegraphics[width=\linewidth]{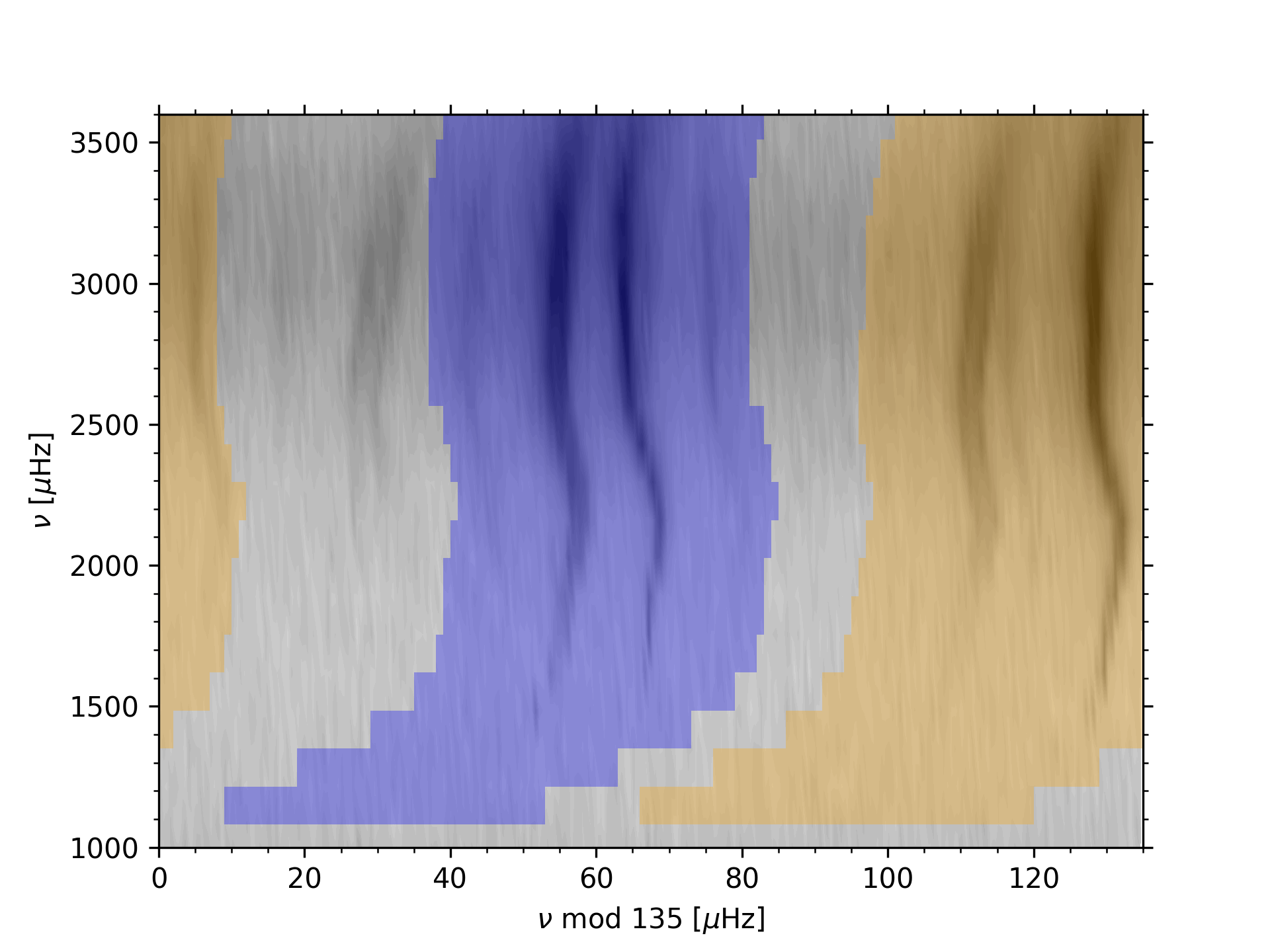}
  \caption{Echelle diagram showing the power in the 45-yr BiSON spectrum in greyscale, with the coloured areas marking the fitting ranges for the $l=2/0$ (blue) and $l=3/1$ (orange) mode pairs. The dark streak to the left of the $l=2/0$ ridge is the $l=4$ mode, which is excluded from the fitting.}
\label{fig:ech}
\end{figure}

We fit the modes in pairs, $(l+2,n-1)/(l,n)$, where $l$ is 0 or 1. To specify the  fitting window we first take the mid-point of the mode pair and then choose the upper and lower limits to be at least {22\,\microHz} above and below this point. 
We also require that the central frequency of each multiplet in the range should be no less than {15\,\microHz} from the end of the range, extending the range if necessary to ensure this. In practice this means that the $l=2/0$ pairs are fitted in a {44\,\microHz} window and the $l=3/1$ pairs, which are more widely separated, in a window of about {50\,--\,55\,\microHz}.
By using this relatively narrow window we avoid interference from the weak $l=4$ mode, which typically appears about {25\,\microHz} below the $l=2$ peak, as shown in Figure~\ref{fig:ech}.

\subsection{Optimization}
\label{sec:optimization}
We use the affine-invariant sampler from the Python \texttt{emcee} package  \citep{2013PASP..125..306F} to perform a Markov Chain Monte Carlo (MCMC) sampling of the model posterior distribution given the observed spectrum. The spectral density is $\chi^2_2$ distributed, and so the log-likelihood function is given by \citep[e.g.][]{Anderson1990}
\begin{equation}
\label{eq:eq3}
    \log\mathcal{L} = -\sum_{i=1}^{N}{\log{M_i} + S_i/M_i},
\end{equation}
where the sum is over the $N$ frequency bins in the range around each mode pair, $M$ is the model and $S$ is the observed spectrum. Note that strictly this equation only applies for data without gaps; introducing gaps lowers the effective resolution of the power spectrum and means that the bins are not independent.

\begin{table}
\caption{Details of the prior distributions for the model parameters. The parameters for the distribution are given as $\mu, \sigma$ for the normal distribution ($\mathcal{N}$) and as the lower and upper bounds for the uniform distribution ($\mathcal{U}$). The values $\mu_{A,l}$ and $\mu_{c}$ are derived from the 
smoothed spectrum, while $\mu_{\nu,l}$ and $\mu_{\Gamma,l}$ are taken from the first-guess table. The value of $\mu_{V,l}$ is 0.54 and 0.38 for $l=2$ and $l=3$ respectively.} 
    \centering
    \begin{tabular}{l|l}
        Parameter &  Prior distribution \\
        \hline
        $\log_{10}{A}$ & $\mathcal{N}(\mu_{A,l}, 5)$ \\
        $b$ & $\mathcal{N}(0, 0.05)$ \\
        $\nu_{l}$ & $\mathcal{N}(\mu_{\nu,l}, 3\mu_{\Gamma,l})$ \\
        $\delta\Omega$ & $\mathcal{N}(0.4, 0.1)*\mathcal{U}(0,\infty)$ \\
        $\log_{10}{\Gamma_{l}}$ & $\mathcal{N}(\mu_{\Gamma,l}, 0.5)$\\
        $\log_{10}{c}$ & $\mathcal{N}(\mu_{c,l}, 5)$ \\
        $V_l$ & $\mathcal{N}(\mu_{V,l}, 0.2\mu_{V,l})*\mathcal{U}(0, \infty)$ \\
        $\log_{10}(\Gamma_{l} / \Gamma_{l+2})$ & $\mathcal{N}(0, 0.2)$ \\
        $b_l-b_{l+2}$ & $\mathcal{N}(0, 0.001)$ \\
        $\nu_{l} - \nu_{l+2}$ & $\mathcal{U}(0, \infty)$ \\
        \hline
    \end{tabular}
    
    \label{tab:priors}
\end{table}

The priors we apply to each of the model parameters are presented in Table \ref{tab:priors}. We use prior distributions based on the Gaussian function for all of the parameters of our model, specified by a centroid value $\mu$ and a width $\sigma$. Hard constraints -- where the log-probability goes to $-\infty$ if the limit is exceeded -- are used to keep the splitting and the visibility factor positive, modifying the underlying Gaussian prior for those parameters. 
The centroids of the prior distributions for the frequency ($\nu$) and width ($\Gamma$) parameters are taken from a first-guess table in which the frequencies are based on the results of \citet{2009MNRAS.396L.100B} and the mode widths are from a smoothed version of earlier historical fits to low-degree modes. For amplitude, line width and background offset terms we vary the logarithm of the parameter, so it is the logarithmic value that is drawn from the distribution. 
 Some additional prior constraints are included in the calculation of the prior probability function in order to keep the widths within a mode pair reasonably close, and 
to ensure that the asymmetry values within a pair are strongly correlated. A further hard constraint on the difference between frequencies in a mode pair is applied to prevent the modes in the pair from swapping places, which can otherwise occasionally happen when the mode width is approaching the separation between modes. 

We use 100 walkers over 2000 steps and discard any walkers with an acceptance fraction below 0.16. While there is no formal criterion for the convergence of an MCMC fit, from visual inspection of the evolution of the parameters during the fit we found that the parameters usually settle around their final value within the first 500 steps.

The full information about the fit result is contained in the posterior distributions of the parameters, and we illustrate some of these below in the results section. We use the median and half of the difference between the 16th and 84th percentile of the marginalized posterior distributions as summary statistics on each of the model parameters when a single uncertainty is needed, while the positive and negative uncertainties shown in Table~\ref{tab:my_label} come from the difference between the median and the 84th and 16th percentiles respectively. In most cases the posteriors for the splitting are Gaussian in form and close to symmetrical, so these are equivalent to 1-sigma errors.

\subsection{Synthetic data}

To test the fitting procedure, we used synthetic BiSON-like ``SolarFLAG'' data prepared as described by \cite{2006MNRAS.369..985C} and \cite{2015MNRAS.454.4120H}. The SolarFLAG data are constructed with $\delta\Omega={400\,{\mathrm{nHz}}}$ and visibility ratios of 0.54 for $l=2, m=0:m=2$ and 0.38 for $l=3, m=1:m=3$, the same values used for the prior distributions in our fitting. We generated 500 independent 11-year realizations, each covering a solar cycle with activity variation based on Cycle 23. We then concatenated these realizations in sets of five, applying the BiSON gap pattern, to simulate the full BiSON history, yielding 100 realizations of a 45-year spectrum. Each of the synthetic spectra was fitted in exactly the same way as the real BiSON one. By averaging the fitting results from many realizations, we can both check that the uncertainties in our results are appropriate and uncover any systematic bias in the results.

\section{Results}

\subsection{Synthetic data test with multiple realizations}
\begin{figure}
\includegraphics[width=\linewidth]{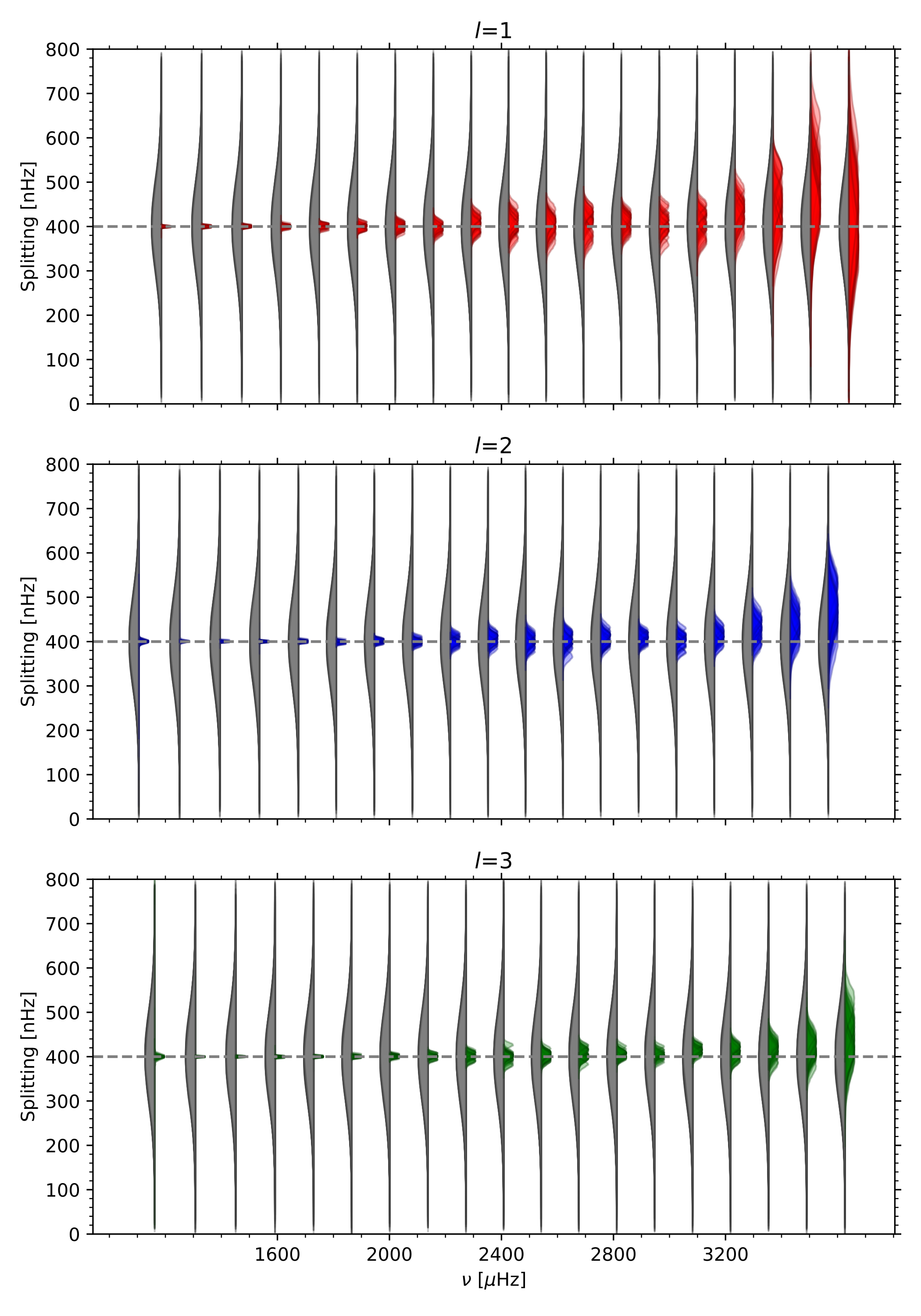}
  \caption{Prior (grey) and posterior (colour) distributions for the rotational splitting parameter. The results of fitting 45-year spectra from 100 realizations of SolarFLAG synethetic data are superimposed. The horizontal dashed line at {400\,nHz} indicates the ``true'' value used to construct the data.}
\label{fig:sf1}
\end{figure}

\begin{figure}
\includegraphics[width=\linewidth]{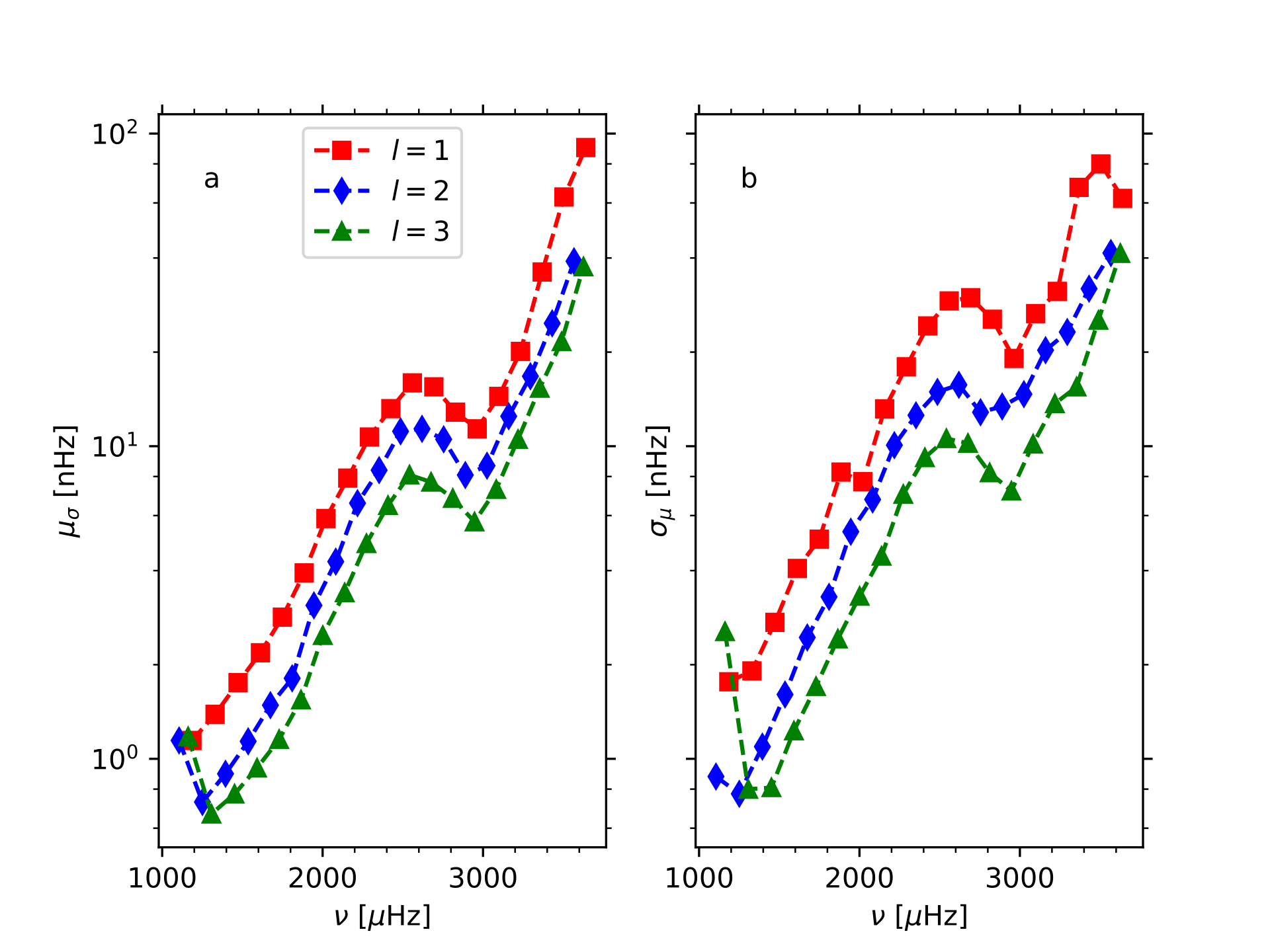}
\includegraphics[width=\linewidth]{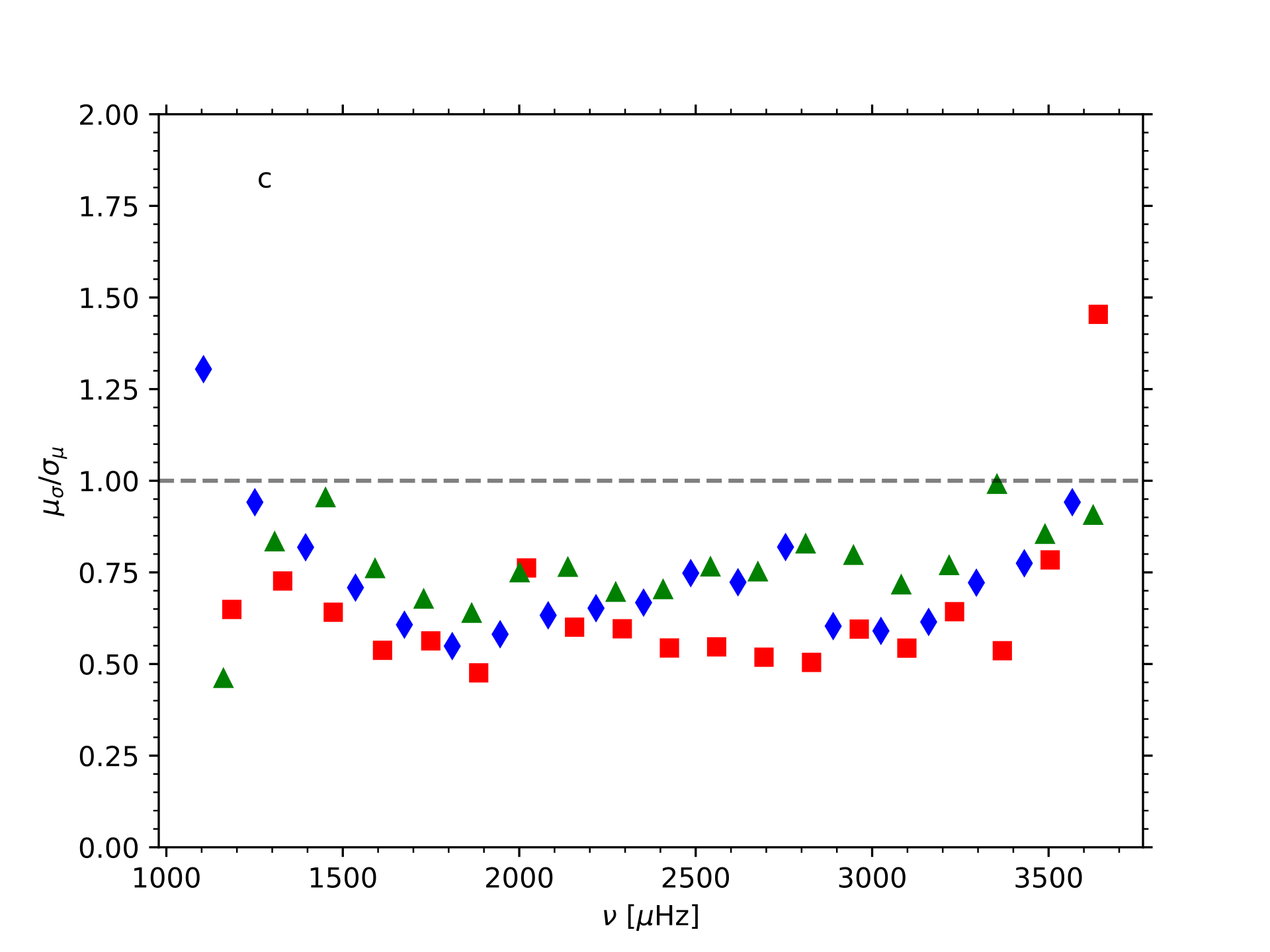}
\caption{The mean width of the posterior distribution of the splitting, $\mu_{\sigma}$ (a), and the standard deviation of the median value, $\sigma_{\mu}$ (b), are shown as a function of frequency for the Monte Carlo test with 100 realizations of the 45-year SolarFLAG spectrum. Panel c shows the ratio of  $\mu_{\sigma}$ to $\sigma_{\mu}$, with the dashed line indicating $y=1$.}
\label{fig:sf2}
\end{figure}

We show in Figure~\ref{fig:sf1} the superimposed posterior distributions of the splitting parameter for fits to our 100 synthetic 45-year spectra. We can see that there is some spread in the centroids of the distributions, increasing with frequency. In Figure~\ref{fig:sf2}a,b we show the mean value $\mu_{\sigma}$ of the 1-sigma error estimated from the distributions and the standard deviation $\sigma_\mu$ of the mean value, and in Figure~\ref{fig:sf2}c we plot the ratio  of $\mu_{\sigma}$ to $\sigma_{\mu}$. We can see that the ratio values cluster below the $y=1$ line, which suggests that the uncertainties from the fitting are underestimated, by about 30 per cent on average. To be more precise, the mean and standard errors of the $\mu_{\sigma}:\sigma_{\mu}$ ratio are $0.64\pm 0.05$ for $l=1$, $0.74\pm 0.04$ for $l=2$, and $0.77 \pm 0.03$ for $l=3$. 
This level of underestimation seems reasonable given that, because of the duty cycle of the data, the frequency bins will not be completely independent and hence 
Equation~\ref{eq:eq3} is not strictly appropriate. In a separate test where we fitted synthetic data with 100 per cent duty cycle there was no such underestimate of the errors.

\begin{figure}
\includegraphics[width=\linewidth]{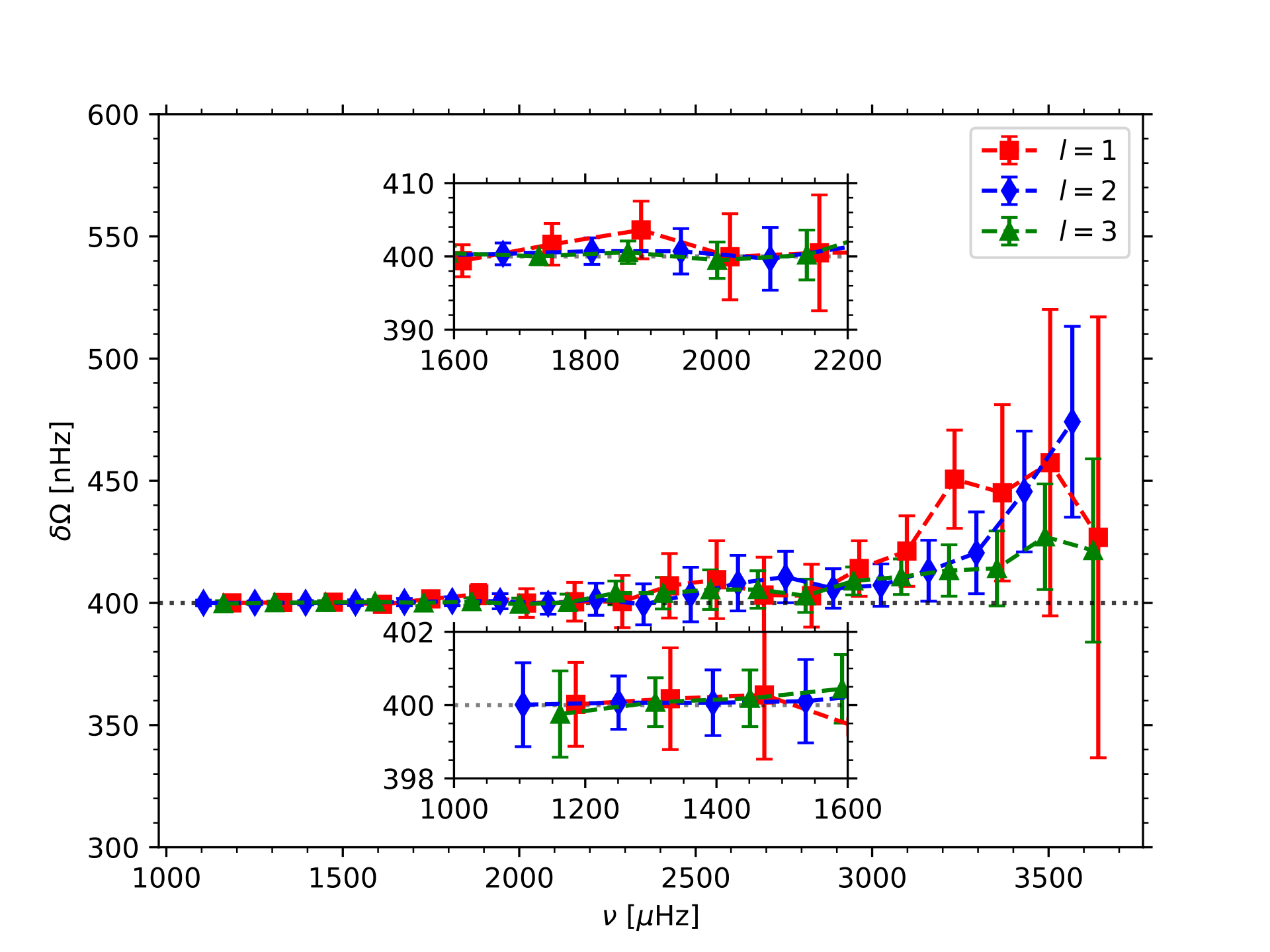}
\caption{The median value of the posterior distribution for the splitting from the Monte Carlo test where 100 realizations of the 45-year SolarFLAG spectrum were fitted using our standard priors. The error bars are taken from the mean width of the posterior distributions, $\mu_{\sigma}$. The horizontal dashed line indicates the ``true'' splitting value of 400\,nHz. The inset plots show the lower-frequency portions of the plot on magnified scales.}
\label{fig:sf3}
\end{figure}

Figure~\ref{fig:sf3} shows the mean centroid $\mu_{\mu}$ of the splitting estimates over our 100 realizations, plotted as a function of frequency with errors taken from $\sigma_{\mu}$. Here we can see that there is an upward bias, increasing with frequency, in the splitting estimates where the mode components are not fully resolved (above $n=14$), but this is generally within the uncertainties and decreases with $l$. In the test with 100 per cent duty cycle this bias is slightly reduced for $l=2$ modes but still present.

\subsection{Synthetic data test for sensitivity to the visibility scale factor}

\begin{figure}
    \includegraphics[width=\linewidth]{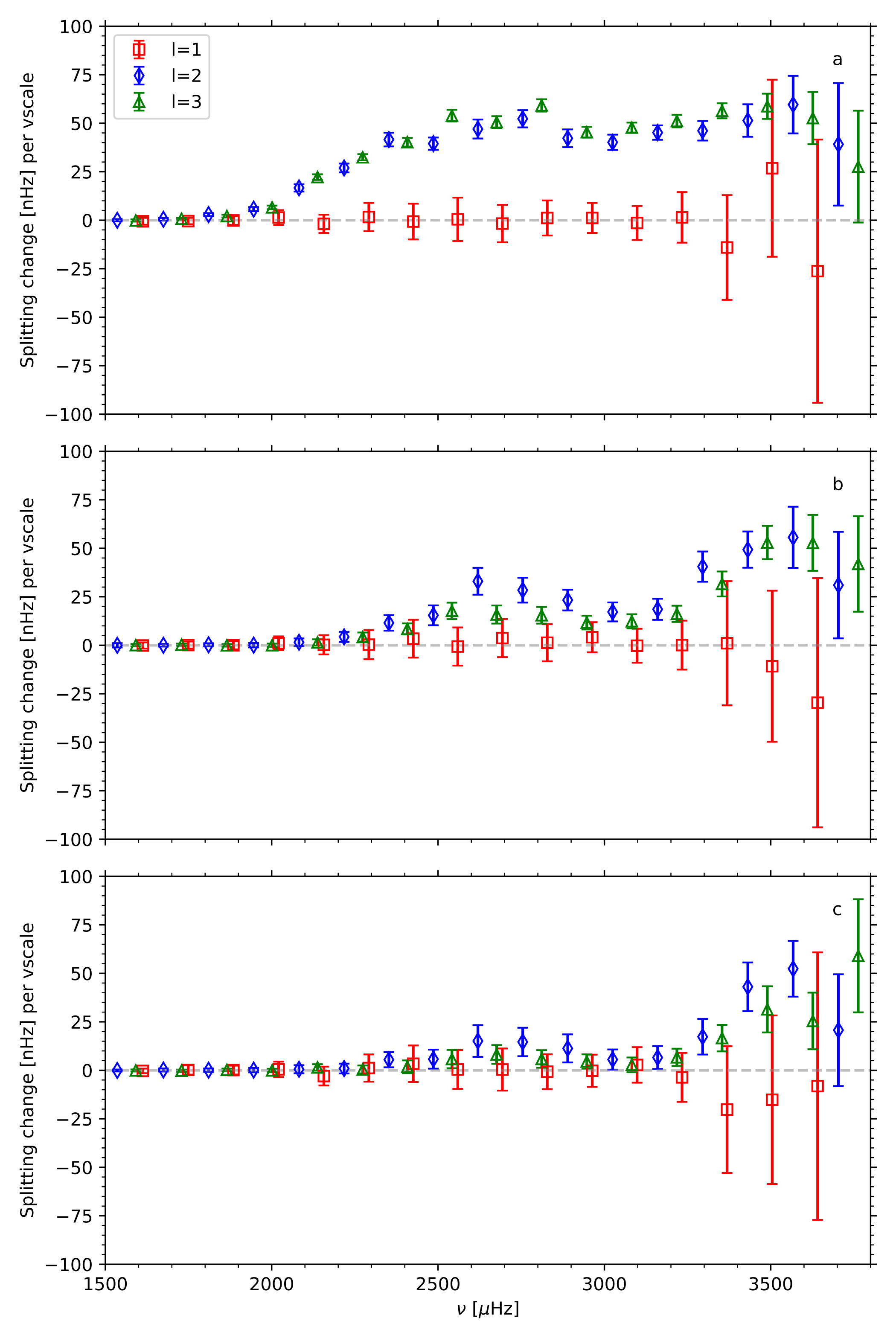}
    \caption{Sensitivity of the splitting estimate to the value of a factor scaling the central value of the prior distribution for the $l=2,m=0:m=2$  and $l=3, m=1:m=3$ ratios when the width of the prior distribution is set at (top) 0.001, (middle), 0.2, and (bottom) 0.4 times the central value, for a single realization of the SolarFLAG spectrum.}
    \label{fig:sensfig}
\end{figure}

It has been shown \citep{2004ESASP.559..356C, 2006MNRAS.369..985C} that choosing the wrong visibility ratios for the $l>1$ components will result in a systematic bias of the splittings. In order to verify that our chosen method of using a constrained free parameter for the ratio will ameliorate this bias, we carried out three sets of tests in which we fitted a single realization of the SolarFLAG spectrum using a central value of the visibility factor prior that was scaled relative to our usual value by factors ranging from 0.7 to 1.3 in steps of 0.15. In the first set of tests, the $\sigma$ of the prior distribution was set at the central value multiplied by 0.001 ; in the second set, it was 0.2 times the central value; and in the final set it was 0.4 times the central value. For each mode within each of the three sets, we then fitted a linear variation to the derived splitting values as a function of the prior central value for the visibility ratio, thus obtaining a value for the sensitivity of the splitting estimate to the central value of the visibility ratio factor. The results are shown in Figure~\ref{fig:sensfig}. This is similar to the test described by \citet{2004ESASP.559..356C}, and the results in the first panel of Figure~\ref{fig:sensfig} are very similar to theirs; above about {2500\,\microHz} the sensitivity is such that a 100 per cent overestimate of the value of the visibility factor would shift the inferred splitting of the $l=2$ and $l=3$ modes upward by about 50\,nHz. For the modes where the components are fully resolved, there is no sensitivity to the visibility factor. In the other two panels, we see that the sensitivity of the $l=2$ and $l=3$ splittings to the visibility factor is substantially reduced by using the wider prior distributions.  Above about 3400\,\microHz, the change of prior has very little effect. Although at the 0.2 prior width the bias is not completely eliminated, we have sufficient confidence in our estimate of the true visibility ratios that we chose to use this prior rather than a wider one. In practice, when fitting the BiSON data, the differences between the splitting results for a 0.4 and 0.2 width of the visibility-factor prior are well within the uncertainties.

\subsection{BiSON data}
We now turn to fitting the 45-year BiSON spectrum.

\begin{figure*}
    \centering

    \includegraphics[width=0.45\linewidth]{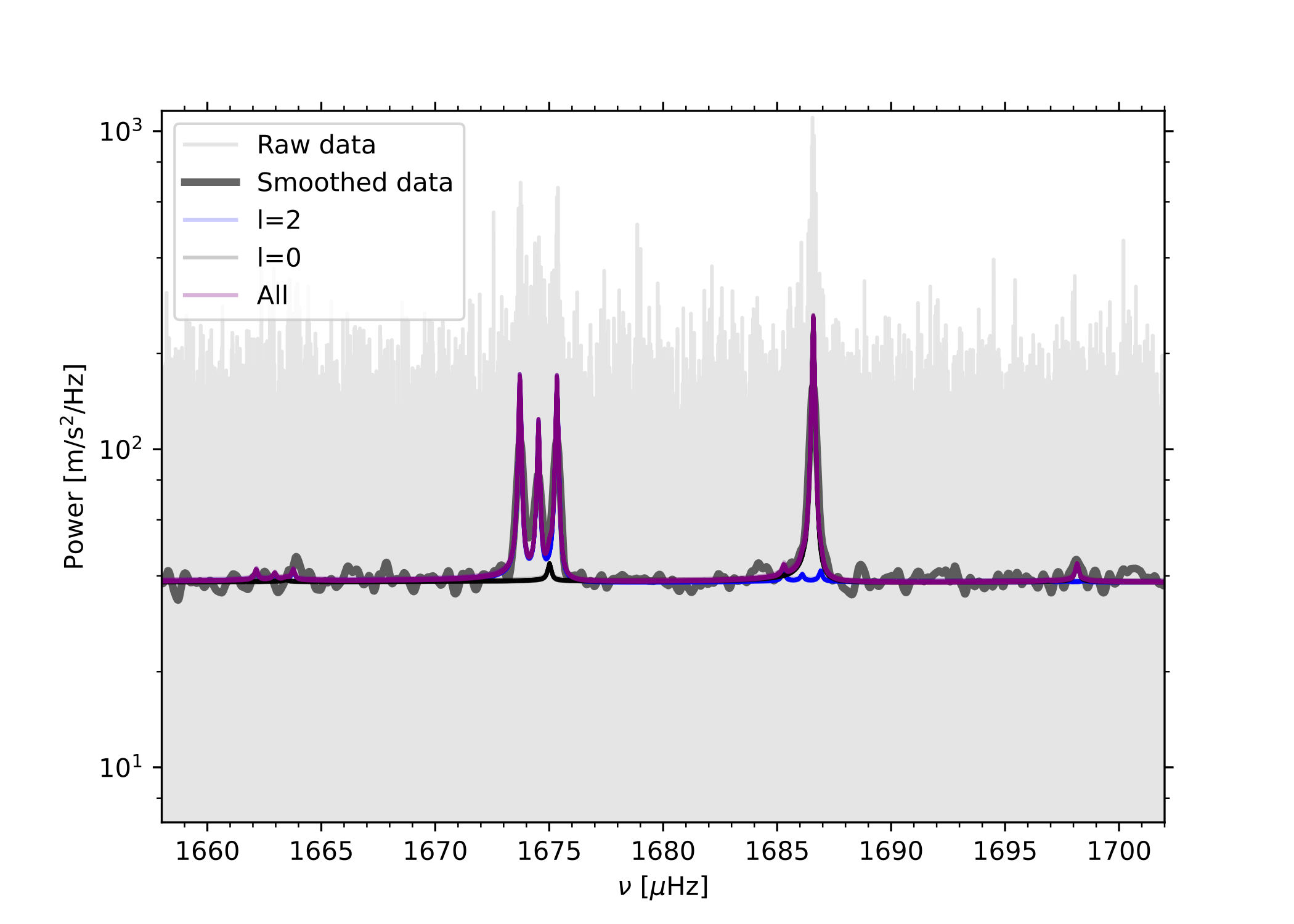}
    \includegraphics[width=0.45\linewidth]{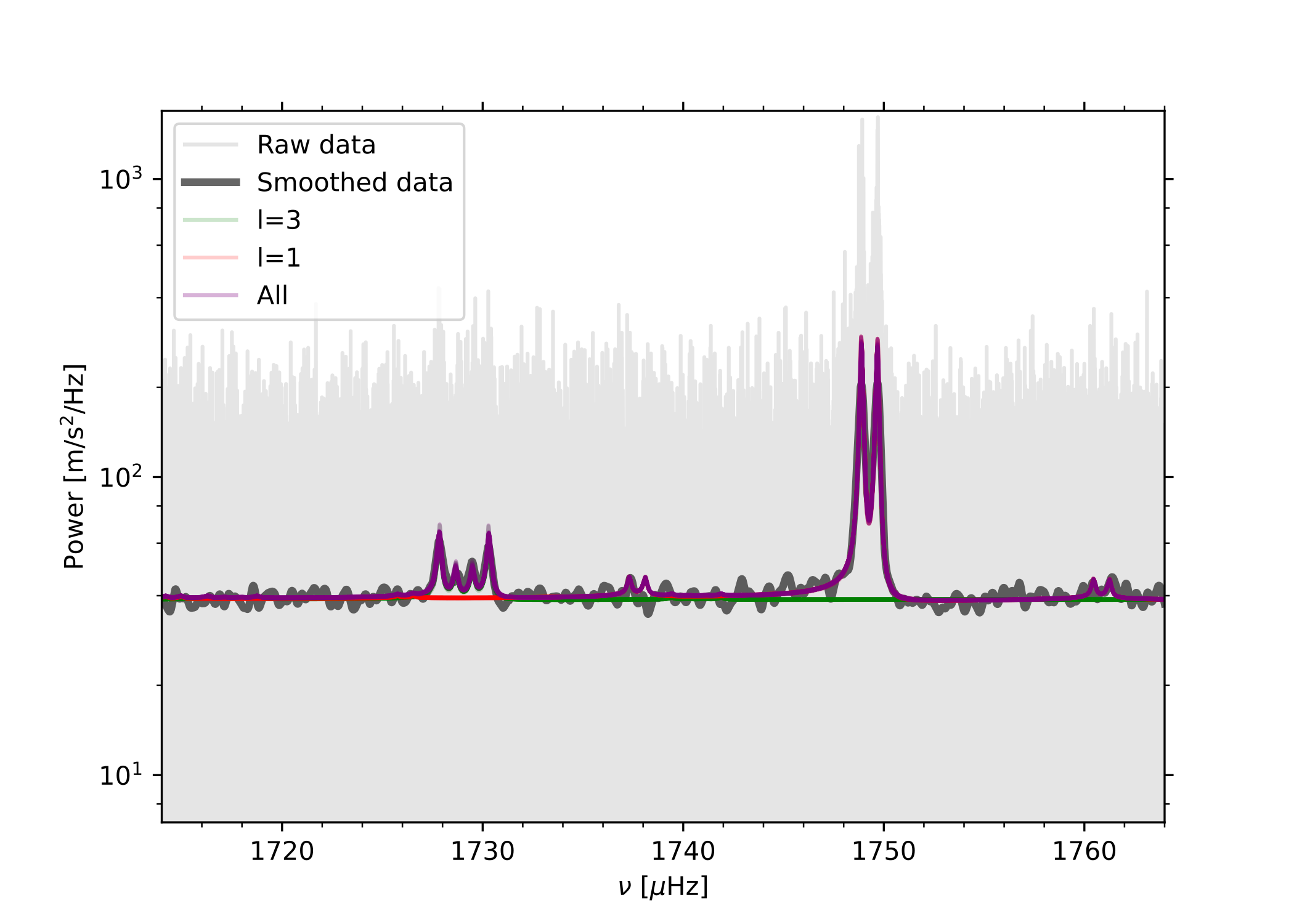}

    \includegraphics[width=0.45\linewidth]{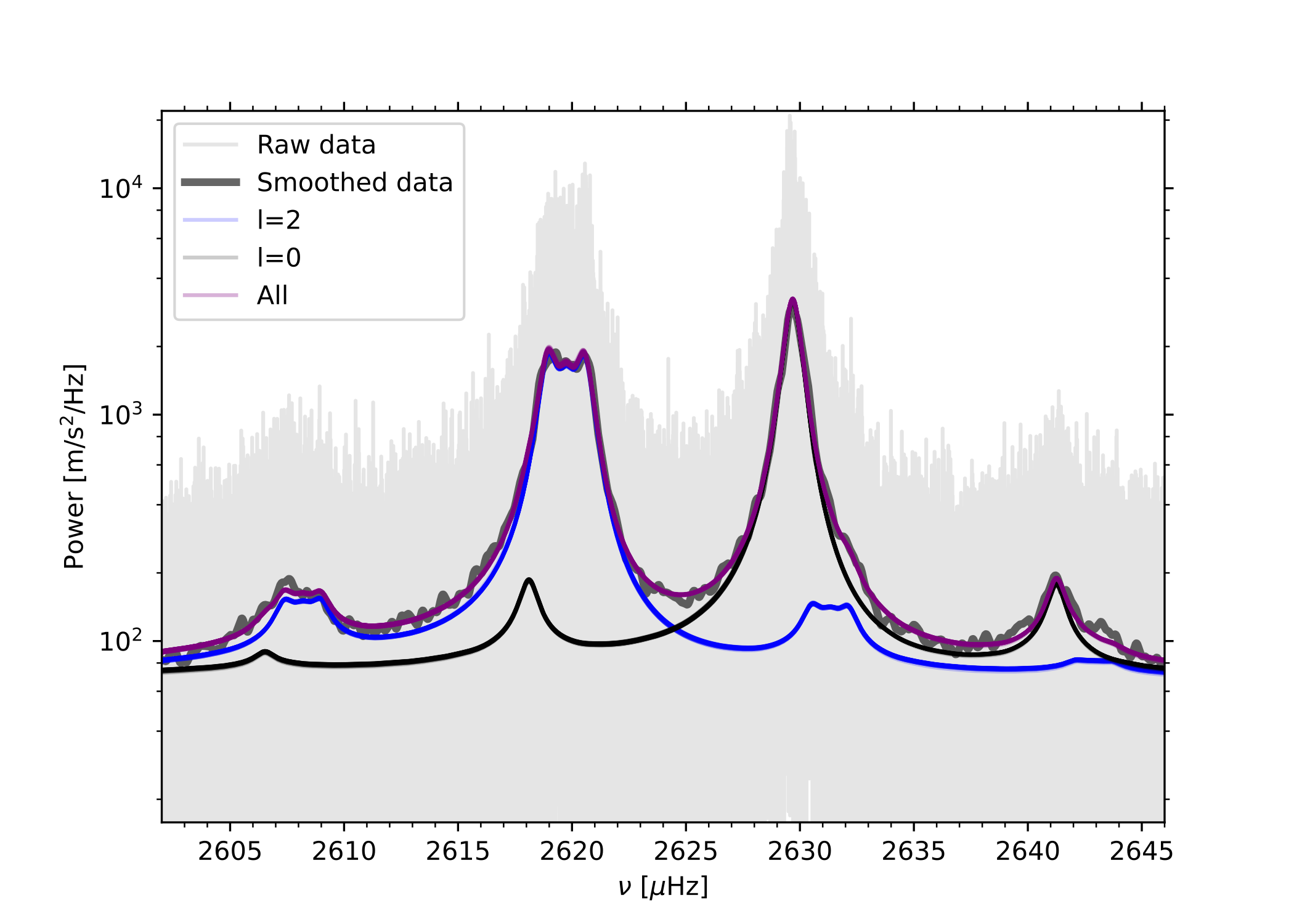}
    \includegraphics[width=0.45\linewidth]{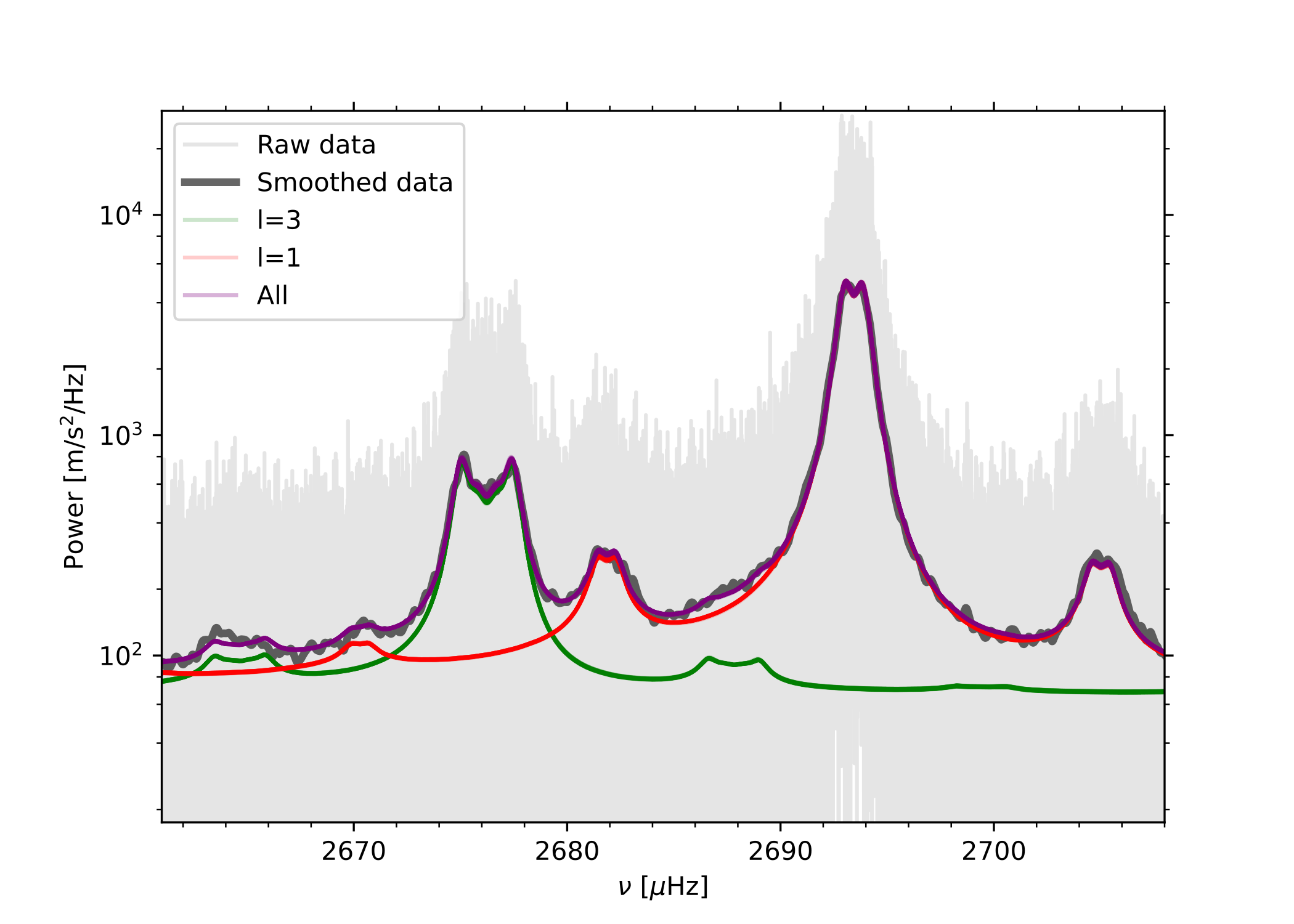}
    
    \caption{Sections of the 45-year BiSON spectrum, showing fits to resolved (top) and partially resolved (bottom) mode pairs with $l=2/0$ (left) and $l=3/1$ (right). A sample of the fitted models derived from the posterior parameter distributions are plotted in purple. The contributions to the fitted model from each mode are shown in different colours as indicated by the legend. 
    The unsmoothed spectrum is shown in light grey; note that it extends beyond the boundaries of the plots.  
    The smoothed spectrum is also shown as the darker grey curve (mostly obscured by the model fits), to illustrate the agreement between the fitted model and the limit spectrum. 
    }
    \label{fig:my_label4}
\end{figure*}

Figure~\ref{fig:my_label4} shows selected mode-pair fits, plotted on a logarithmic scale. The smoothed input power spectrum is plotted, but it is almost obscured by the fitted model, giving a visual indication that the fit is working well. In the Appendix, we show the corresponding ``corner plots'' for these fits.

\begin{figure}
    \centering
    \includegraphics[width=\linewidth]{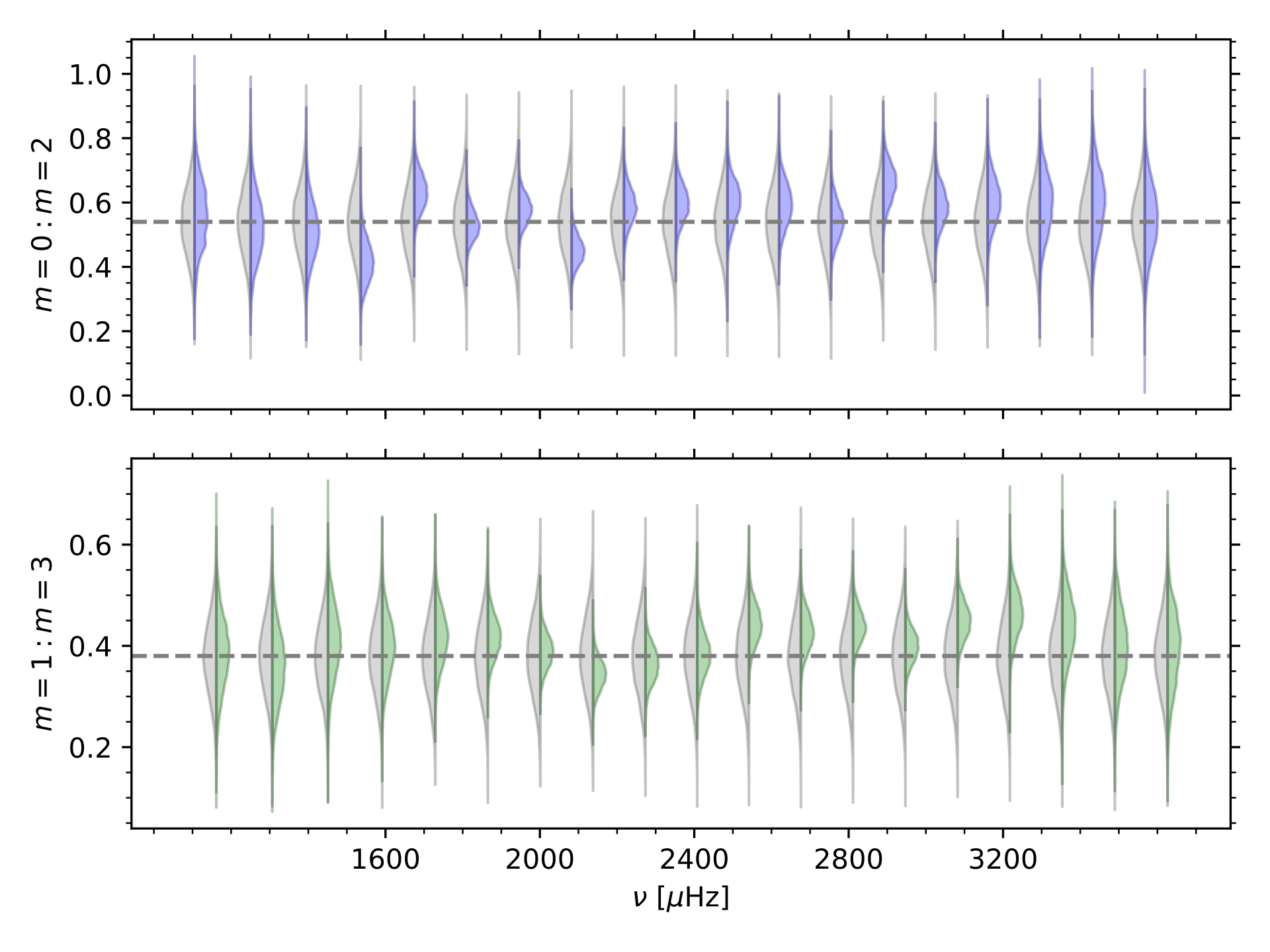}
    \caption{Prior (grey) and posterior (coloured) distributions of the visibility ratio parameter for $l=2$ (top) and $l=3$ (bottom), for the 45-year BiSON spectrum. The grey horizontal dashed lines show the centroid for the prior distribution.}
    \label{fig:my_label6}
\end{figure}

In Figure~\ref{fig:my_label6} we show the prior and posterior distributions for the visibility ratios for the $l=2$ and $l=3$ modes as a function of frequency. For most of the modes these are dominated by the prior; at low frequencies this is because the signal-to-noise ratio of the inner components is low, while at high frequencies the components are not resolved. For a small number of modes between about {1700\,--\,2200\,\microHz} the visibility parameter appears to be somewhat constrained by the fit, and the posterior distributions suggest that the choice of prior was appropriate. We
consider it reasonable to use the same prior for all frequencies. In recent resolved-Sun work \citep[e.g.][]{2015SoPh..290.3221L,2023FrASS...931313K} a term dependent on the radial order is included in the calculation of the leakage matrix (to which our visibility factors correspond) to account for the horizontal displacement of the modes, but according to \cite{2015SoPh..290.3221L} it is reasonable to neglect this for high-order modes. 

\begin{figure}
    \centering
    \includegraphics[width=\linewidth]{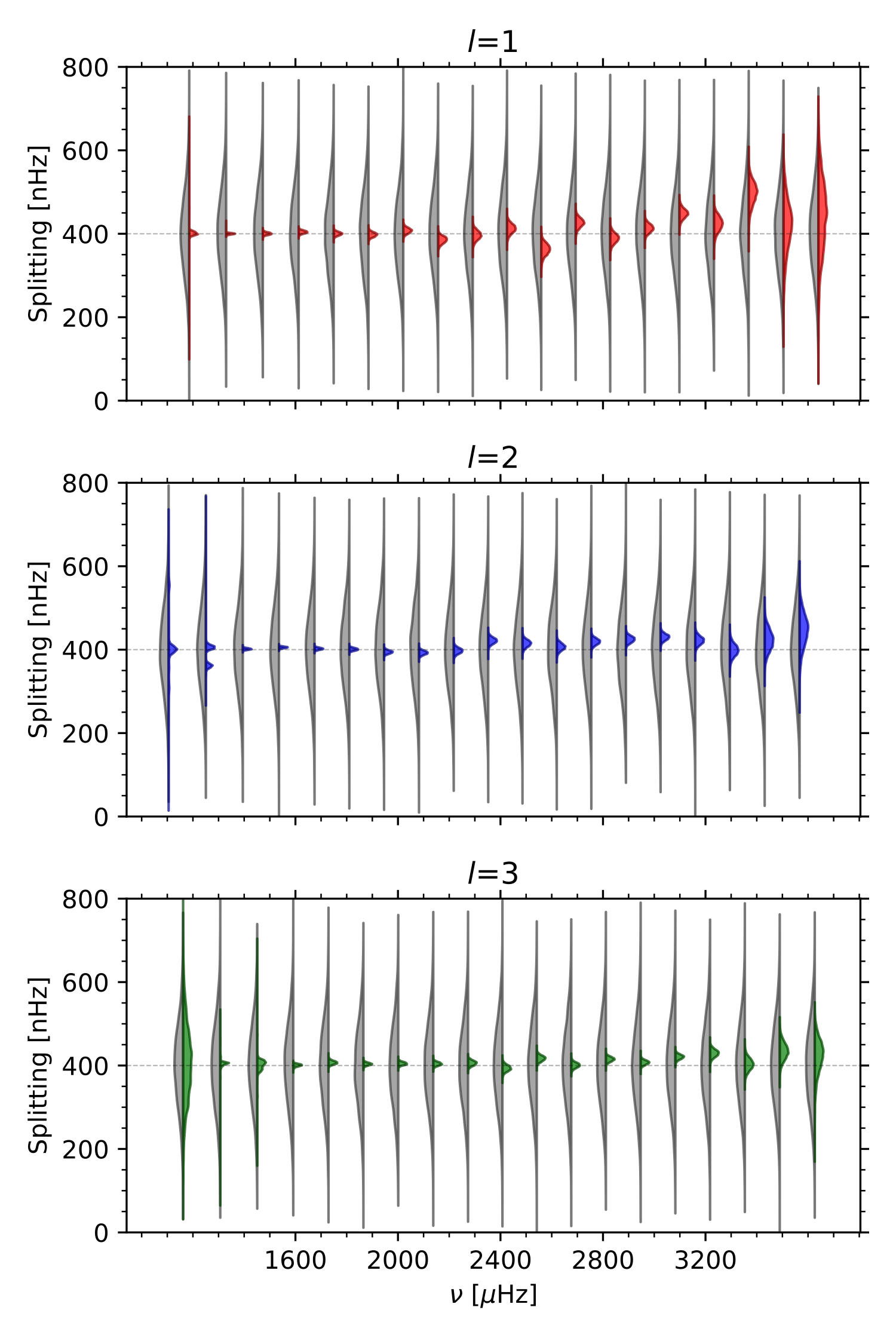}
    \caption{Prior (grey) and posterior (coloured) distributions for $l=1$, $l=2$, and $l=3$ splittings from the 45-year BiSON spectrum, as a function of frequency. The grey horizontal lines show the median of the prior distribution.}
    \label{fig:my_label7}
\end{figure}

Figure ~\ref{fig:my_label7} shows the prior and posterior distributions for the mode splitting, and Table~\ref{tab:my_label} gives the median values and $\pm 1\sigma$ uncertainties taken from the distributions. We note that the distributions for the two highest orders of $l=1$ (which are unresolved) and the lowest order of $l=3$ (which is lost in the background noise) are prior-dominated and are not included in the table. There is a spurious second peak in the distribution for the $l=2, n=7$ mode at {1250\,\microHz}, reflected in a very asymmetric uncertainty estimate for this mode, which is not included in the table. Again, this is a case where the signal-to-noise ratio of the mode is poor and the fit may be affected by a noise spike. Otherwise, the splittings appear to be well constrained and the posterior distributions are close to Gaussian. In Table~\ref{tab:my_labeld} we show the splittings and uncertainties after adjusting for the bias and underestimation of uncertainties identified by using the synthetic data: the uncertainties have been scaled by the $\mu_{\sigma}:\sigma_{\mu}$ ratio from the Monte Carlo experiment and the difference between the mean splitting estimates from the synthetic data and the true 400\,nHz value has been subtracted from the splitting estimates. 

We can estimate a mean splitting over several modes by combining samples drawn from the posterior distribution for each. After applying the adjustments used in Table~\ref{tab:my_labeld}, for $8 \leq n \leq 15$ we obtain values of 399.3$_{-9.6}^{+6.4}$\,nHz for $l=1$, 400.3$_{-7.4}^{+4.2}$\,nHz for $l=2$, and 403.5$_{-5.0}^{+5.4}$\,nHz for $l=3$, while for $16 \leq n \leq 23$ the values are 
402.6$_{-36.0}^{+30.1}$\,nHz for $l=1$, 411.6$_{-12.3}^{+11.5}$\,nHz for $l=2$, and 407.6$_{-12.0}^{+9.5}$\,nHz for $l=3$. All of these values are consistent with the 400\,nHz value used to construct the SolarFLAG data.

\begin{figure}
    \centering
\includegraphics[width=\linewidth]{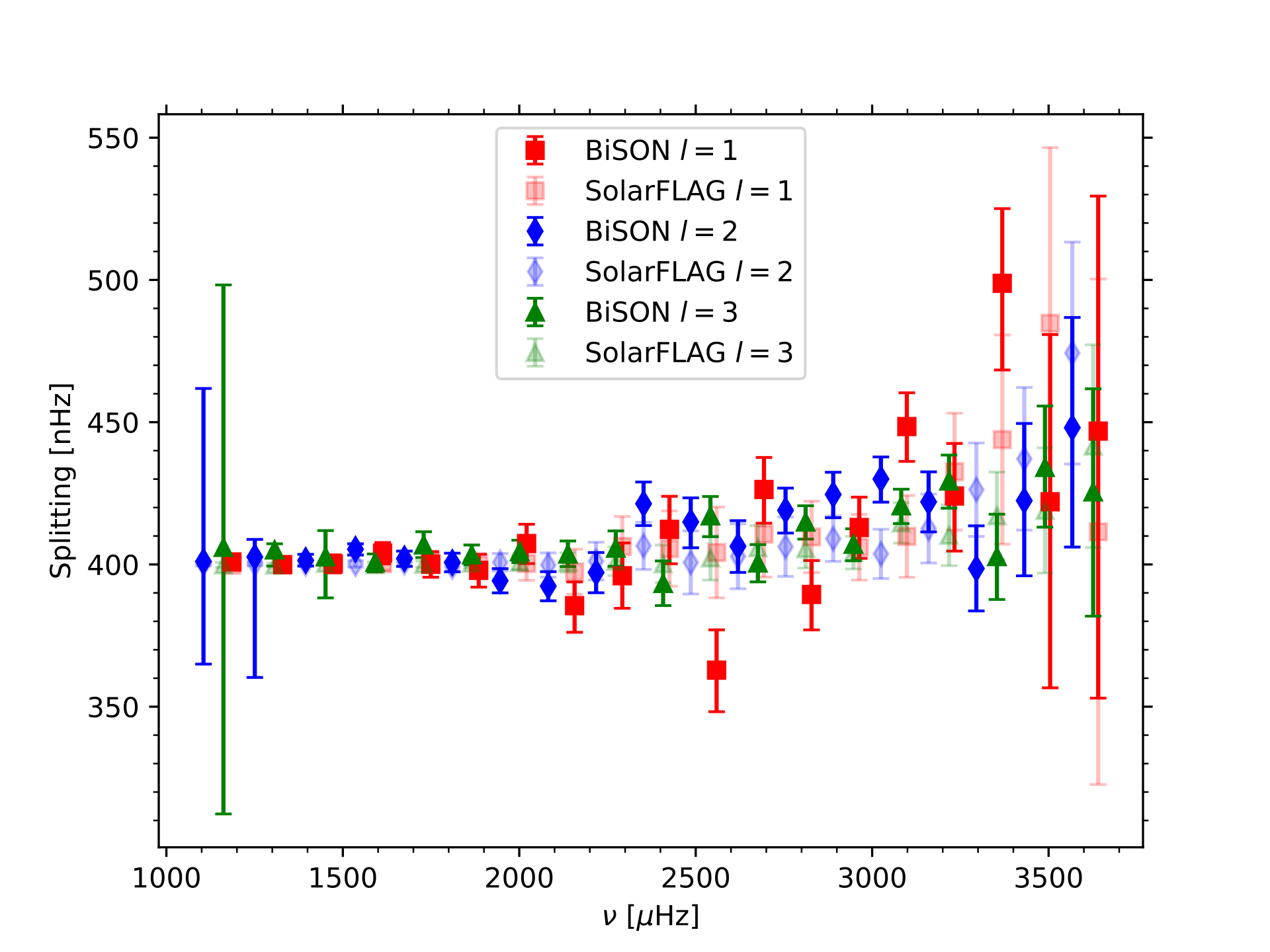}
    \caption{Splittings from the 45-year BiSON spectrum in error-bar format, as a function of frequency. The synthetic SolarFLAG results from Figure~\ref{fig:sf3} are overlaid in fainter shades.}
    \label{fig:my_label8}
\end{figure}

Finally, in Figure~\ref{fig:my_label8} we show the splitting values as a function of frequency in error-bar form, overlaid on the results from Figure~\ref{fig:sf3}. The trends at frequencies above {2000\,\microHz} appear generally consistent, with a few notable outliers. The low observed splitting value for the $l=1, n=17$ mode at {2559\,\microHz} is particularly striking and seems to be robust, as it occurs even in fitting with different choices for the priors, but it is most likely a result of the stochastic nature of the mode excitation rather than an intrinsic feature of the underlying rotation profile. The $l=1$ splittings in general show some scatter around the synthetic average values, but there is no obvious systematic trend.

\begin{table*}
  \caption{Median value and upper and lower 1-$\sigma$ widths (derived from the 16th and 84th percentiles) for the marginalized posterior distributions of the 
 synodic splitting parameter, for the 45-yr BiSON spectrum. Note that the quoted uncertainties are underestimated by about 30 per cent. Splittings for modes above $n=15$ are biased upwards because the components are not fully resolved, and these splittings should not be used in inversions.}

\begin{tabular}{rrllrllrll}
\hline
   $n$ &   $l$ & ${\nu}$ [$\mu$Hz]   & Splitting [nHz]   &   $l$ & ${\nu}$ [$\mu$Hz]        & Splitting [nHz]       &   $l$ & ${\nu}$ [$\mu$Hz]   & Splitting [nHz]   \\
\hline

     7 &     1 & 1185.6              & 400.9$_{-2.7}^{+2.5}$ &     2 & 1250.6 & \ldots   &     3 & 1306.7              & 405.1$_{-5.6}^{+2.2}$ \\
     8 &     1 & 1329.6              & 399.9$_{-1.8}^{+1.7}$ &     2 & 1394.7 & 401.4$_{-2.0}^{+2.1}$ &     3 & 1451.0              & 403$_{- 14}^{+  9}$   \\
     9 &     1 & 1472.8              & 400.2$_{-3.1}^{+3.2}$ &     2 & 1535.9 & 405.3$_{-2.0}^{+1.9}$ &     3 & 1591.5              & 400.7$_{-3.4}^{+2.9}$ \\
    10 &     1 & 1612.7              & 404$_{-4}^{+4}$   &     2 & 1674.5 & 402.0$_{-2.7}^{+2.7}$ &     3 & 1729.1              & 407$_{-  4}^{+  5}$   \\
    11 &     1 & 1749.3              & 400$_{-5}^{+4}$   &     2 & 1810.3 & 401$_{-  3}^{+  3}$   &     3 & 1865.3              & 403$_{-  4}^{+  3}$   \\
    12 &     1 & 1885.1              & 398$_{-6}^{+6}$   &     2 & 1945.8 & 394$_{-  4}^{+  4}$   &     3 & 2001.2              & 404$_{-  4}^{+  4}$   \\
    13 &     1 & 2020.8              & 407$_{-7}^{+7}$   &     2 & 2082.1 & 392$_{-  5}^{+  5}$   &     3 & 2137.8              & 404$_{-  4}^{+  5}$   \\
    14 &     1 & 2156.8              & 385$_{-  9}^{+  8}$   &     2 & 2217.7 & 397$_{-  7}^{+  7}$   &     3 & 2273.5              & 406$_{-  6}^{+  6}$   \\
    15 &     1 & 2292.0              & 396$_{- 11}^{+ 12}$   &     2 & 2352.3 & 421$_{-  8}^{+  8}$   &     3 & 2407.7              & 393$_{-  8}^{+  8}$   \\
    16 &     1 & 2425.6              & 412$_{- 12}^{+ 12}$   &     2 & 2485.9 & 415$_{-  9}^{+  8}$   &     3 & 2541.7              & 417$_{-  7}^{+  7}$   \\
    17 &     1 & 2559.2              & 363$_{- 15}^{+ 14}$   &     2 & 2619.8 & 406$_{-  9}^{+  9}$   &     3 & 2676.2              & 401$_{-  7}^{+  6}$   \\
    18 &     1 & 2693.4              & 426$_{- 12}^{+ 11}$   &     2 & 2754.5 & 419$_{-  8}^{+  8}$   &     3 & 2811.4              & 415$_{-  6}^{+  6}$   \\
    19 &     1 & 2828.2              & 389$_{- 12}^{+ 12}$   &     2 & 2889.7 & 425$_{-  8}^{+  8}$   &     3 & 2947.0              & 407$_{-  6}^{+  5}$   \\
    20 &     1 & 2963.4              & 413$_{- 11}^{+ 11}$   &     2 & 3024.8 & 430$_{-  8}^{+  8}$   &     3 & 3082.4              & 421$_{-  6}^{+  6}$   \\
    21 &     1 & 3098.3              & 448$_{- 12}^{+ 12}$   &     2 & 3160.0 & 422$_{- 11}^{+ 11}$   &     3 & 3217.8              & 429$_{- 10}^{+  9}$   \\
    22 &     1 & 3233.3              & 424$_{- 19}^{+ 18}$   &     2 & 3295.2 & 399$_{- 15}^{+ 15}$   &     3 & 3353.6              & 403$_{- 15}^{+ 15}$   \\
    23 &     1 & 3368.7              & 499$_{- 30}^{+ 26}$   &     2 & 3430.9 & 422$_{- 26}^{+ 27}$   &     3 & 3489.7              & 434$_{- 21}^{+ 22}$   \\
    24 &     1 & 3504.4              & 420$_{- 70}^{+ 60}$   &     2 & 3567.1 & 450$_{- 40}^{+ 40}$   &     3 & 3626.3              & 430$_{- 40}^{+ 40}$   \\
 
\hline
\end{tabular}

 \label{tab:my_label}
\end{table*}

\begin{table*}
\caption{Adjusted values of synodic splitting and errors for the 45-yr BiSON spectrum. The uncertainties have been scaled by the $\sigma_{\mu}:\mu_{\sigma}$ ratio from the Monte Carlo experiment, and the difference between the mean splitting and the true 400\,nHz value in the Monte Carlo experiment  has been subtracted from the splitting value, for each mode.}
\begin{tabular}{rrllrllrll}
\hline
   $n$ &   $l$ & ${\nu}$ [$\mu$Hz]   & Splitting [nHz]       &   $l$ & ${\nu}$ [$\mu$Hz]   & Splitting [nHz]       &   $l$ & ${\nu}$ [$\mu$Hz]   & Splitting [nHz]       \\
\hline
 
     7 &     1 & 1185.6              & 401$_{-  4}^{+  4}$   &     2 & 1250.6              & \ldots   &     3 & 1306.7              & 405.0$_{-6.8}^{+2.6}$ \\
     8 &     1 & 1329.6              & 399.7$_{-2.5}^{+2.3}$ &     2 & 1394.7              & 401.4$_{-2.4}^{+2.6}$ &     3 & 1451.0              & 403$_{- 15}^{+ 10}$   \\
     9 &     1 & 1472.8              & 400$_{-  5}^{+  5}$   &     2 & 1535.9              & 405.2$_{-2.9}^{+2.7}$ &     3 & 1591.5              & 400$_{-  4}^{+  4}$   \\
    10 &     1 & 1612.7              & 405$_{-  7}^{+  7}$   &     2 & 1674.5              & 402$_{-  4}^{+  4}$   &     3 & 1729.1              & 407$_{-  7}^{+  7}$   \\
    11 &     1 & 1749.3              & 398$_{-  8}^{+  8}$   &     2 & 1810.3              & 400$_{-  6}^{+  6}$   &     3 & 1865.3              & 403$_{-  6}^{+  5}$   \\
    12 &     1 & 1885.1              & 394$_{- 12}^{+ 12}$   &     2 & 1945.8              & 394$_{-  7}^{+  7}$   &     3 & 2001.2              & 405$_{-  5}^{+  5}$   \\
    13 &     1 & 2020.8              & 407$_{-  9}^{+  9}$   &     2 & 2082.1              & 393$_{-  8}^{+  8}$   &     3 & 2137.8              & 403$_{-  6}^{+  6}$   \\
    14 &     1 & 2156.8              & 385$_{- 16}^{+ 14}$   &     2 & 2217.7              & 396$_{- 11}^{+ 11}$   &     3 & 2273.6              & 402$_{-  9}^{+  9}$   \\
    15 &     1 & 2292.0              & 395$_{- 19}^{+ 19}$   &     2 & 2352.3              & 422$_{- 12}^{+ 11}$   &     3 & 2407.7              & 389$_{- 11}^{+ 11}$   \\
    16 &     1 & 2425.6              & 405$_{- 22}^{+ 21}$   &     2 & 2485.9              & 411$_{- 12}^{+ 11}$   &     3 & 2541.7              & 412$_{- 10}^{+  9}$   \\
    17 &     1 & 2559.2              & 353$_{- 27}^{+ 26}$   &     2 & 2619.8              & 398$_{- 13}^{+ 12}$   &     3 & 2676.2              & 395$_{-  9}^{+  9}$   \\
    18 &     1 & 2693.4              & 423$_{- 23}^{+ 22}$   &     2 & 2754.5              & 408$_{- 10}^{+ 10}$   &     3 & 2811.4              & 412$_{-  7}^{+  7}$   \\
    19 &     1 & 2828.2              & 386$_{- 25}^{+ 24}$   &     2 & 2889.7              & 419$_{- 14}^{+ 13}$   &     3 & 2947.1              & 398$_{-  7}^{+  7}$   \\
    20 &     1 & 2963.4              & 399$_{- 18}^{+ 18}$   &     2 & 3024.8              & 423$_{- 14}^{+ 13}$   &     3 & 3082.4              & 410$_{-  9}^{+  8}$   \\
    21 &     1 & 3098.3              & 427$_{- 23}^{+ 22}$   &     2 & 3160.0              & 409$_{- 17}^{+ 17}$   &     3 & 3217.8              & 416$_{- 12}^{+ 12}$   \\
    22 &     1 & 3233.3              & 373$_{- 30}^{+ 29}$   &     2 & 3295.2              & 378$_{- 21}^{+ 21}$   &     3 & 3353.6              & 389$_{- 15}^{+ 15}$   \\
    23 &     1 & 3368.7              & 450$_{- 60}^{+ 50}$   &     2 & 3430.9              & 380$_{- 30}^{+ 40}$   &     3 & 3489.7              & 407$_{- 25}^{+ 25}$   \\
    24 &     1 & 3504.4              & 360$_{- 80}^{+ 70}$   &     2 & 3567.1              & 370$_{- 40}^{+ 40}$   &     3 & 3626.3              & 400$_{- 50}^{+ 40}$   \\
  
\hline
\end{tabular}

\label{tab:my_labeld}
\end{table*}

\section{Discussion and Conclusions}

We have derived rotational splitting parameters for modes of $l=1, 2$ and $3$ from a BiSON spectrum spanning 16,425 days from 1976 December 31\,--\,2021 December 20. Our algorithm assumes a single splitting parameter per mode, with the visibility ratio for the inner components of $l=2$ and $l=3$ modes as a variable parameter. The estimated value for the visibility ratio is dominated by its prior except for the few orders where the inner components are both well resolved and have good signal-to-noise ratios. We present the posterior distributions from our MCMC fits as well as the centroid estimates and their uncertainties. It should be noted that the $l=3$ splitting values do not include a differential rotation term and are not strictly equivalent to the first-order term of a polynomial expansion of frequency as a function of $m$.

We have presented the results for modes where the splittings are not resolved, for the sake of completeness. However, the Monte Carlo experiments we have performed with simulated data suggest that the splitting estimates for these modes ($n>14$) are systematically biased upwards and should not be used for inversions, at least not without appropriate correction. This bias appears in $l=1$ modes as well as the higher-degree modes, so we believe it is intrinsic to the problem of fitting unresolved peaks and not related to the modeling of the visibility function of the modes nor to the neglect of various higher-order and yearly effects that are included neither in our fits nor in the synthetic data. The upward trend in splitting with frequency in the fits to BiSON data appears to be consistent with that from SolarFLAG data where the splitting is constant at 400\,nHz, and therefore these results do not suggest that there is any real upward trend in the splitting with frequency.

Our Monte Carlo experiments also reveal that the uncertainties from the fitting are underestimated by about 30 per cent. We believe this is due to the reduction in the effective resolution of the power spectrum because the overall duty cycle is relatively low due to sparse observations early in the time series. The uncertainties should be corrected accordingly when using the splittings together with data from other sources, and we have provided a table of data and uncertainties adjusted based on the results of our Monte Carlo experiments.

We note that the BiSON observations have a variable duty cycle, being sparser in the early years, and hence the measurements from the 45-yr spectrum are not a uniform average over those years but more heavily weighted towards the post-1992 epoch.

In future work we hope to compare splittings estimates from a shorter subset of the BiSON data with those from space-based Sun-as-a-star instruments such as GOLF and resolved-Sun instruments such as the Global Oscillations Network Group (GONG), the Michelson Doppler Imager (MDI) and the Helioseismic and Magnetic Imager (HMI). We also plan to refine the fitting algorithm to provide definitive values for the mode frequencies.

\section*{Acknowledgments}
We acknowledge the support of the UK Science and Technology Facilities Council (STFC) through
grant ST/V000500/1.

The computations described in this paper were performed using the University of Birmingham's BlueBEAR HPC service, which provides a High Performance Computing service to the University's research community. See \url{http://www.birmingham.ac.uk/bear} for more details. 

This research made use of 
NASA's Astrophysics Data System.

We would like to thank all those who have been associated with BiSON over the years. We thank Dr. Annelies Mortier for useful suggestions.

We thank the anonymous referee for helpful comments.

\section*{Software}
Below we include additional software used in this work which has not explicitly been mentioned above.
\begin{itemize}
    \item Python \citet{python1995}
    \item AstroPy \citep[][]{astropy:2013,astropy:2018}
    \item Matplotlib \citet{Hunter:2007}
    \item NumPy \citet{harris2020array}
    \item SciPy \citet{2020SciPy-NMeth}
    \item corner \cite{corner}
\end{itemize}

\section*{Data Availability}

 The BiSON time series analysed here is available at \url{http://bison.ph.bham.ac.uk/opendata}.

The SolarFLAG time series are available on reasonable application to the authors.

\bibliography{bison}

\appendix
\section{Corner plots}

\begin{figure*}
    \includegraphics[width=\linewidth]{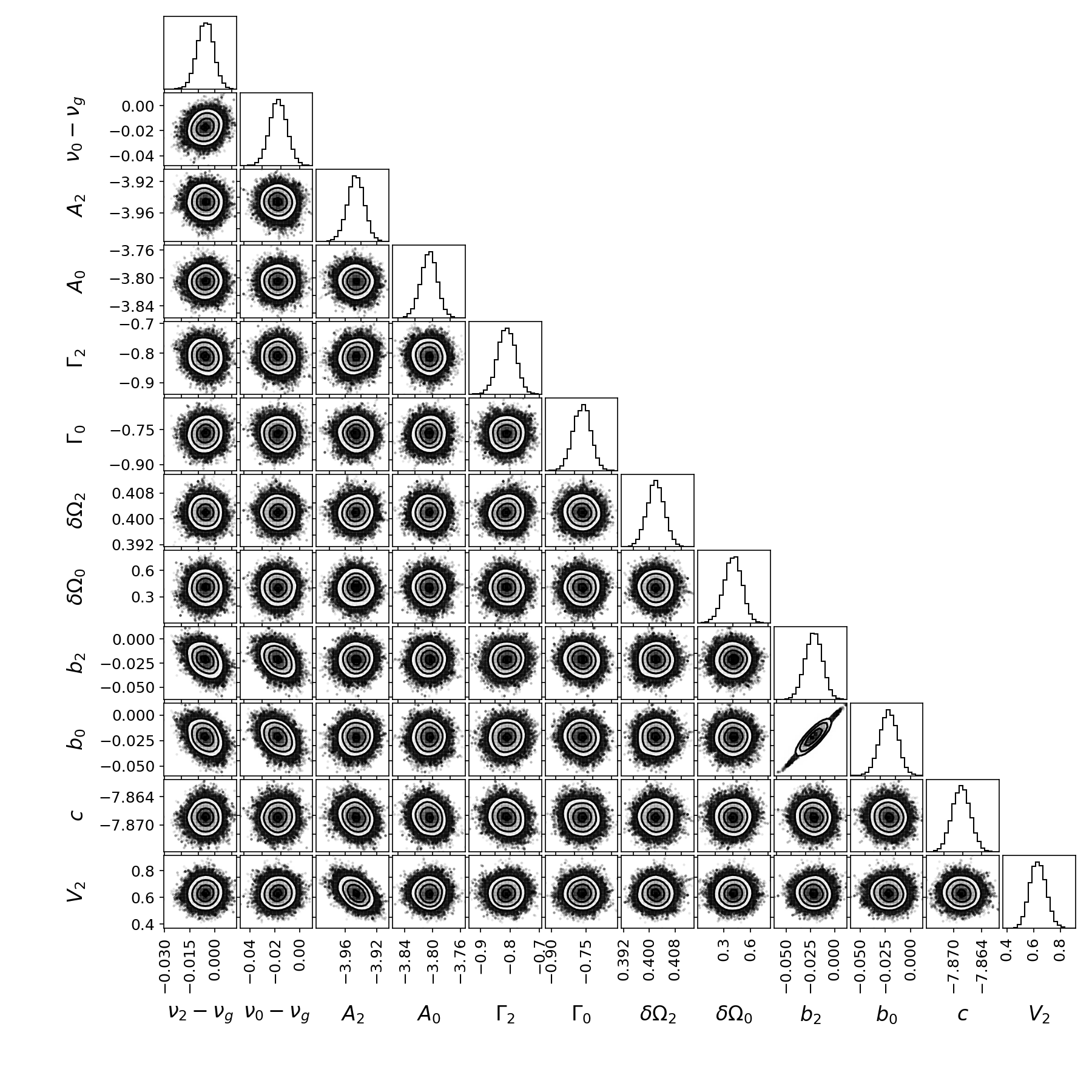}
    \caption{Corner plot for fit to the $l=2,n=10/l=0,n=11$ mode pair.}
    \label{fig:corn1}
\end{figure*}

\begin{figure*}
    \includegraphics[width=\linewidth]{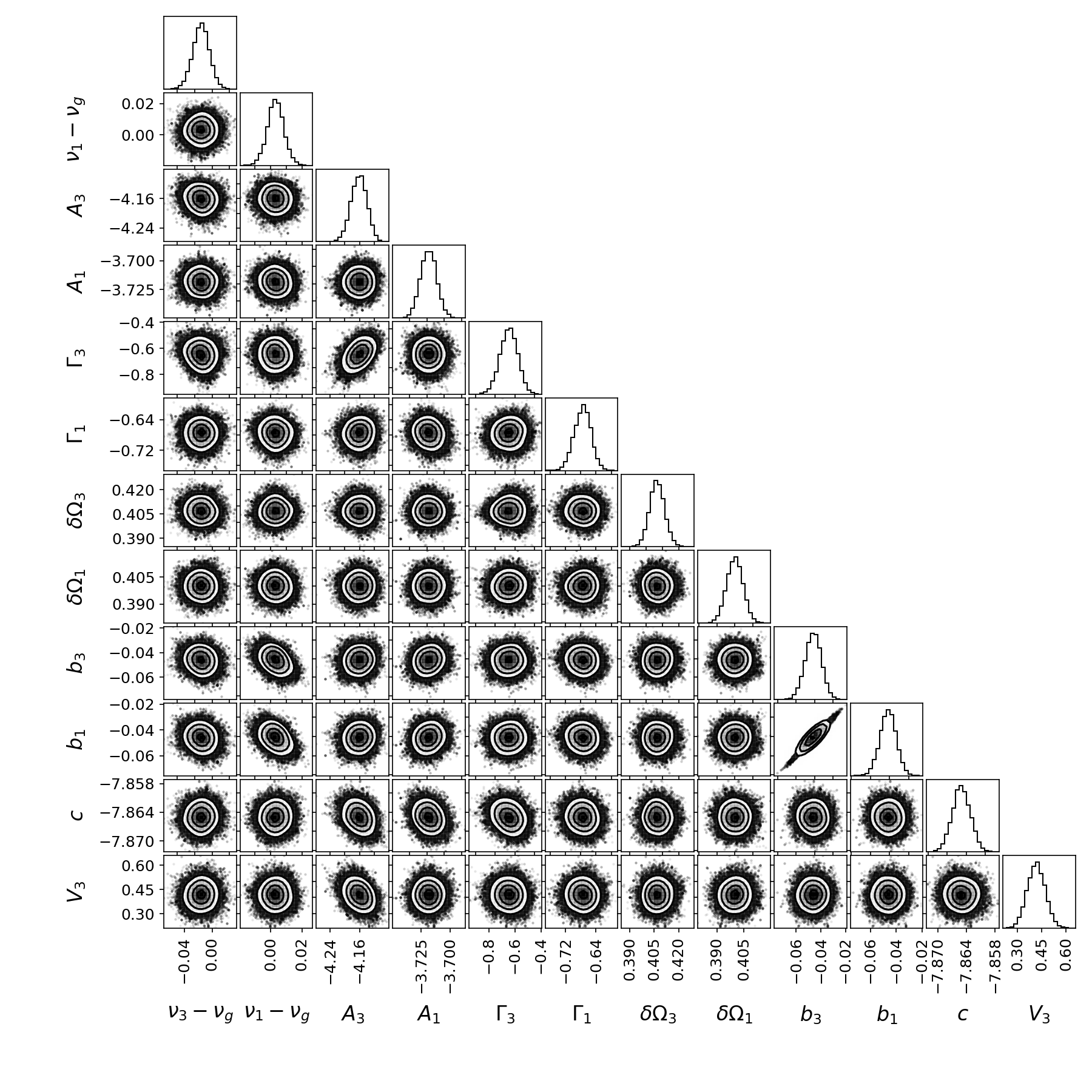}
    \caption{Corner plot for fit to the $l=3,n=10/l=1,n=11$ mode pair.}
    \label{fig:corn2}
\end{figure*}

\begin{figure*}
    \includegraphics[width=\linewidth]{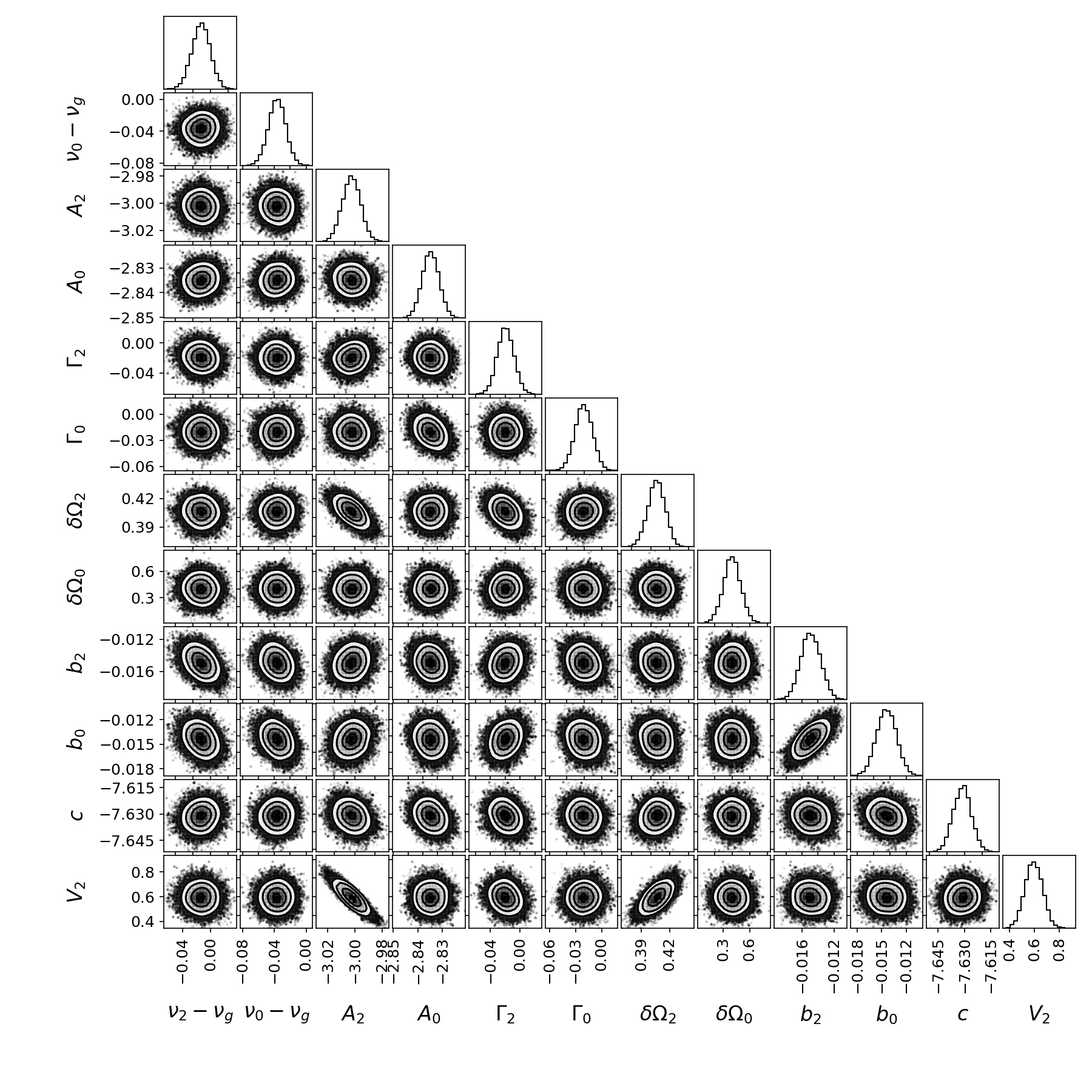}
\caption{Corner plot for fit to the $l=2,n=17/l=0,n=18$ mode pair.}
\label{fig:corn3}
\end{figure*}

\begin{figure*}
    \includegraphics[width=\linewidth]{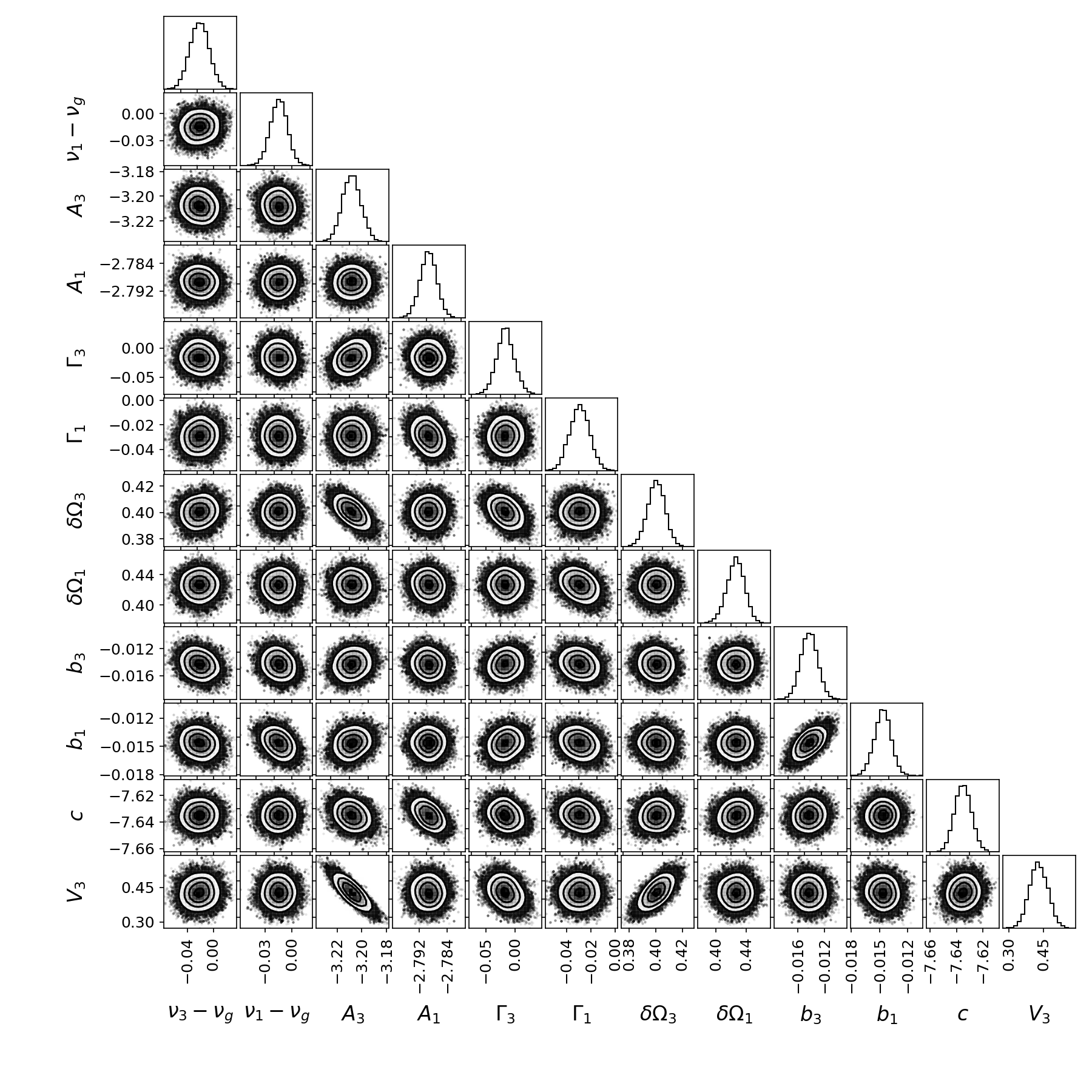}
\caption{Corner plot for fit to the $l=3,n=17/l=1,n=18$ mode pair.}
\label{fig:corn4}
\end{figure*}

Figures~\ref{fig:corn1} to ~\ref{fig:corn4} show the ``corner plots'' for sample fits to the BiSON spectrum. These are the fits that are shown in Figure~\ref{fig:my_label4}. We can see that there some correlations among the parameters, especially for the higher-frequency case. There tends to be a positive correlation between the visibility scale parameter and the $l$=2 or 3 splitting, which is not unexpected. The $l=3$ amplitude is somewhat negatively correlated with the visibility scale parameter, and hence it is also anticorrelated with the $l=3$ splitting. None of the correlations we see are of particular concern, given that our final results are taken from the marginalized posterior distributions.

\label{lastpage}
\end{document}